\documentclass[a4paper,11pt]{article}

\usepackage{jheppub} 

\usepackage[T1]{fontenc} 

\usepackage{bm}
\usepackage{latexsym}
\usepackage{dcolumn}
\usepackage{amsmath,amsfonts,amssymb, amsthm}
\usepackage{graphicx,epsfig}
\usepackage{environ}
\usepackage{mathtools}
\usepackage{braket}
\usepackage{float}
\usepackage{mathrsfs}
\usepackage{amsmath}
\usepackage{fancyhdr}
\usepackage{subcaption}
\usepackage{hyperref}
\usepackage{graphicx,epstopdf}
\usepackage{capt-of}
\hypersetup{
	colorlinks   = true, 
	urlcolor     = blue, 
	linkcolor    = blue, 
	citecolor   = red 
}

\def\ta{\tau} 
\def\e{\eta}
\def\s{\sigma} 
\def\o{\omega}

\def\a{\alpha}
\def\b{\beta}

\def\d{\delta}
\def\s{\sigma} 
\def\O{\Omega}
\def\ph{\phi}

\def\p{\partial}
\def\f{\frac}

\def\be{\begin{equation}}
\def\ee{\end{equation}}
\def\beq{\begin{eqnarray}}
\def\eeq{\end{eqnarray}}
\def\p{\partial}

\title{Conformal Vacuum and Fluctuation-Dissipation in de-Sitter Universe and Black Hole Spacetimes}
\author{Ashmita Das$^a$\footnote {email:\color{blue} ashmita@iitg.ac.in, ashmita.phy@gmail.com}~,}
\author{Surojit Dalui$^a$\footnote {email:\color{blue} suroj176121013@iitg.ernet.in}~,}
\author{Chandramouli Chowdhury$^b$\footnote {email:\color{blue} chandramouli.chowdhury.icts.res.in, chandramouli.chowdhury@gmail.com}~,}
\author{Bibhas Ranjan Majhi$^a$\footnote {email:\color{blue} bibhas.majhi@iitg.ac.in}~,}
\affiliation{$^a$Department of Physics, Indian Institute of Technology Guwahati,\\ 
Guwahati 781039, Assam, India.}
\affiliation{$^b$International Centre for Theoretical Sciences, Bengaluru, North Karnataka 560089, India.}

\abstract{In the studies of quantum field theory in curved spacetime, the ambiguous concept of vacuum state and the particle content is a long-standing debatable aspect. So far it is well known to us that in the background of the curved spacetime, some privileged class of observers detect particle production in the suitably chosen vacuum states of the quantum matter fields. In this work we aim to study the characteristics behaviour of these produced particles in the background of the de-Sitter (dS) Friedmann-Lama\^{i}tre-Robertson-Walker (FLRW) Universe (both for $(1+1)$ and $(3+1)$ dimensions) and $(1+1)$-dimensional Schwarzschild black hole (BH) spacetime, from the point of view of the respective privileged class of observers. Here the analysis is confined to the observers who perceive particle excitations in the conformal vacuum. We consider some test particles in the thermal bath of the produced particles and calculate the correlation function of the fluctuation of the random force as exerted by the produced quanta on the test particles. We obtain that the correlation function abides by the fluctuation-dissipation theorem, which in turn signifies that the test particles execute Brownian-like motion in the thermal bath of the produced quanta. }
\begin{document}
\maketitle
\flushbottom
\noindent

\section{Introduction}
In quantum field theory (QFT), the proper description of physical vacuum is an interesting and long-standing question. Probably the most straightforward answer to this question is, ``the vacuum is a no particle state'' \cite{Birrell}. However, since 1970, consistent development of the studies related to the observer dependence in QFT is suggesting that the quantum measuring process, plays the most crucial role to describe a no/many-particle state \cite{Birrell}. 
For instance, an inertial measuring device, i.e., a detector/observer records no particle excitations in the Minkowski vacuum of a quantum field whereas a uniformly accelerated detector records particle excitations from the same vacuum state \cite{Fulling:1972md,
Unruh:1973}. 
Therefore the state of the motion of the detecting device is a significant factor in describing a physical vacuum and is subject of consistent interest in theoretical as well as experimental physics, since decades. However, for Minkowski spacetime, there is a conventional description of the vacuum state, where the vacuum is defined with respect to all the inertial measuring devices. On the other hand in the curved spacetime, there are many possibilities of the vacuum states corresponding to a quantum field, and therefore one cannot define a unique vacuum state of the field \cite{Birrell,Wald:2009uh}. 
 Therefore, physicists have almost agreed that for a proper formulation of QFT in curved spacetime, we need to discard the search for the notion of a {\it unique} vacuum state and build up the formalism on the basis of the {\it specific} choice of vacuum states. After selecting the specific vacuum state, one can indeed obtain the Hilbert space representation of the states, and subsequently define the Fock space and the field operator, with respect to the choice of the vacuum state. The different choices of vacuum state produce different theoretical outcomes, and hence in the studies of QFT in the curved background, the choice of the vacuum state is of utter significance \cite{Birrell,Wald:2009uh}.

The search for a unique description of vacuum is associated with the fact that the notion of particle content in QFT is an observer-dependent concept. The state of motion of the observer/detecting device, which is employed to detect the particle excitations, plays the most crucial role to define the idea of particle content in a state impeccably. Even in the Minkowski spacetime, without the precise knowledge of the detecting device, the concept of particle content in a state is not accurately defined \cite{Fulling:1972md,Unruh:1973,Birrell}. 
Therefore, with time, the studies of the detection of particles with respect to the different conditions of detecting device, became much more popular than searching for the proper notion of particle concept in QFT. Several pioneering attempts have been accomplished in literatures \cite{Fulling:1972md,Unruh:1973,Parker:1968,Parker:1969,Parker:1971,Davies:1974,Hawking:1974sw,Bunch:1978,Gibbons:1977} to examine the phenomena of particle production from a specific choice of vacuum state, with respect to the particular observer, in curved and as well as flat spacetime. The important aspect of these works is the presence of the ``privileged class of observers'', which can detect particle production from the certain ``vacuum states''. 
In this context some recent developments can be found in \cite{Nikolic:2001ky,Barcelo:2006np,Pereira:2009kv,MartinMartinez:2012sg,Firouzjaee:2015bqa,Markkanen:2015xuw,deHaro:2015hdp,Paliathanasis:2016dhu,Celani:2016cwm}.

With the progress of these investigations of particle production, some authors have taken further initiatives to address a spontaneous question in this context that, ``how these produced particles behave, with respect to the particular observer who is detecting their existence?'' \cite{Sriramkumar, Bibhas_fdt,Rigopoulos:2016oko}. The authors of \cite{Sriramkumar}, have investigated the random motion of a moving mirror, which is immersed in a thermal bath of massless scalar particles. They have calculated the mean radiation reaction force on the mirror as exerted by the scalar particles and the correlation function of the fluctuations in the force about the mean value. From the obtained correlation function, they have established that in the {\it non-relativistic limit} the motion of the moving mirror is dictated by the fluctuation-dissipation theorem (FDT) \cite{Kubo}. Subsequently, in \cite{Bibhas_fdt}, it has been shown that, from the perspective of a uniformly accelerated observer the test particles which are released in the thermal bath of produced particles in the Minkowski vacuum of a massless scalar field, execute Brownian-like motion due to the random force exerted on it by the produced particles. They also have shown that the correlation function of the fluctuations in the random force follow FDT.  

Until now the quest for understanding the nature of the produced quanta have been largely restricted to the background of the flat spacetime, whereas limited analyses have been carried out to comprehend the nature of the motion of the produced quanta in curved spacetime. Although, in curved spacetime, the phenomena of observer-dependent particle production (in a certain choice of vacuum state of a quantum field) is a well-recognised phenomena \cite{Unruh:1973,Parker:1968,Parker:1969,Parker:1971,Davies:1974,Hawking:1974sw,Bunch:1978,Gibbons:1977}.  For example, it is established in literature that in the background of the spatially flat de Sitter FLRW spacetime a comoving observer detects quanta of radiation from the conformal vacuum of the scalar field. Similarly in the $(1+1)$-dimensional Schwarzschild spacetime a static Schwarzschild observer records particle creation in the Unruh and Kruskal vacuum states (see \cite{Birrell} for further references). In other words, one can comment that the privileged class of observers will conclude the existence of thermal spectrum from certain vacuum states, as they are capable to detect the quanta of the field from the corresponding vacuum state.

In this manuscript, we consider the phenomena of particle production in a particular vacuum state in curved spacetime and aim to analyse the behaviour of the produced particles from the perspective of the privileged class of observers. We choose two important classes of curved backgrounds: one is homogenous, isotropic and exponentially inflationary Universe (both $(1+1)$ and $(3+1)$-dimensional), represented by the de-Sitter FLRW metric and static blackhole (BH) spacetime, represented by ($1+1$)-dimensional Schwarzschild spacetime. Note that all of these spacetimes are conformally flat. As a quantum matter field, we consider the simplest of all, i.e., a massless scalar field, minimally coupled with the curvature scalar of the background spacetime. 
 
For certain cases, particle production (in a specific choice of vacuum state) can be intuitively interpreted (by a particular class of observers) as if the produced particles are immersed in a thermal bath of a specific temperature. This temperature comes out to be the function of the parameters of the corresponding curved spacetime.  
Now, to study the behaviour of the produced quanta of the field in the curved spacetime, we drop some test particles in the thermal bath of the produced particles. In principle, one can find the force exerted by all the produced particles on the test particles. Subsequently, determining the exerted force on the test particles, one can proceed to solve the force equations \cite{Bibhas_fdt}. This would naturally lead to solving a large number of differential equations due to the plethora of produced particles in the thermal bath. In these circumstances, the idea of using statistical calculations may turn out to be helpful \cite{Sriramkumar,Bibhas_fdt}.

Thus, at first, we define the force as exerted by the produced quanta of the scalar field (which are basically the produced particles in the thermal bath) on the test particle. Due to a large number of the produced particles, one can indeed realise that the force exerted by them on the test particle is {\it random} by nature. In a phenomenological fashion, we identify this random force in terms of the stress-energy tensor of the quantum scalar field. Subsequently, we compute the fluctuations in the random force and find the force-force correlation function. This correlation function is evaluated with respect to the specific vacuum state (where they were produced) and measured by the particular class of observer. Our purpose of finding the force-force correlation function is to understand the impacts of the produced scalar quanta on the test particles at different points in the thermal bath. This allows one to draw some inference about the dynamics of the produced particles. 

 With this setup, we obtain that in curved spacetime, the correlation function of the fluctuations in the random force follows the FDT. Therefore to the eyes of the observer (who observes the particle production), the test particle released in the thermal bath will execute Brownian-like motion, due to its interaction with the produced particles. As the correlation function of the fluctuations in the random force obeys the FDT, the produced particles themselves start to execute Brownian-like motion in the thermal bath. We also calculate the dissipative coefficient corresponding to all the cases which we have discussed in this manuscript. We mention that the functional form of the dissipative force corresponding to these systems is not known to us. Therefore finding the dissipative coefficient of the system, in turn, may help us to build up the Langevin equation for these kinds of setup, which we will discuss elsewhere. In this regard, we mention that in the case of BH, Candelas and Sciama \cite {Candelas:1977zz} showed that the two-point correlation function for the gravitational shear satisfies fluctuation-dissipation theorem. Another work \cite{Rigopoulos:2016oko} can be brought in this context where the author has generalised Starobinsky's seminal results on stochastic inflation by implementing the effective action of scalar field fluctuations with wavelengths larger than the de Sitter curvature radius. As a consequence, the infrared dynamics of a light in de Sitter spacetime can be interpreted as Brownian motion in a medium with a specific de Sitter temperature whereas we shall see that our present picture is completely different from this.

We organise our paper as follows. In Section \ref{vacuum}, we introduce the concepts of conformal vacuum states and the observers for both the curved spacetimes and review the phenomena of particle production in curved background briefly. Subsequently, in Section \ref{methodology}, we develop the methodology of our work, where we define the force and the correlation function of the fluctuations of the force. In the next section, we implement the methodology to find the correlation function of the fluctuation of the random force. Then we analyse the property of the obtained correlation function corresponding to the dS FLRW, Schwarzschild BH spacetimes in Section \ref{fd_property}.  Section \ref{dissipation} contains the analysis to find the dissipative coefficients corresponding to the curved spacetimes of our interest. Finally, we conclude our work in Section \ref{conclusion}. Five appendices are provided for the ease of our readers, where we explicitly calculate the renormalised expectation value of the components of the stress-energy tensor. These analyses are undoubtedly significant in the context of the particle production in a specific vacuum state of a quantum field in curved spacetime. Also, an appendix has been added at the end to summarise our various notations. Remember that the signature of our spacetimes is $(+,-,-,-)$ through out the paper.
 
 \section{Conformal vacuum and the concept of particle for conformally Minkowski spacetimes}\label{vacuum}
 This section will be dedicated to discussing the notion of conformal vacuum and particle production for a class of conformally flat spacetimes. We shall indicate the observers (frames) which are relevant to the present context as we go along. Although such a discussion exists in literature, we briefly review the relevant portions here. We shall find that this section will act as the backbone of our main goal and thereby initialise the flow of the paper. Moreover, it will make the paper self-sufficient to a reader. For a more elaborate discussion of these topics, we refer the reader to \cite{Birrell}.

\subsection{Conformal vacuum and Green functions}\label{description_frw}
It is well discussed in literature \cite{Birrell,Wald:2009uh,Parker:1968,Parker:1969,Parker:1971,MartinMartinez:2012sg} that, if a curved spacetime possesses some geometrical symmetries, a particular vacuum state and mode solutions of field equations \& particle states can be defined in that curved spacetime. In this context, one of the promising candidates is a curved spacetime with conformal symmetry where in particular, the spacetime is conformally Minkowski (conformally flat). Our aim here is to understand the definition of {\it vacuum}  and how the Green functions corresponding to the curved spacetime can be evaluated by knowing those of Minkowski spacetime. Note that this analysis is valid for massless scalar fields only.

 We start with a massless scalar field $\ph(x)$, non-minimally coupled to the scalar curvature of a general curved spacetime, as $\xi \mathcal{R}\phi^2$, where, $\xi$ is the coupling strength of the scalar field to the Ricci scalar, $\mathcal{R}$. $\mathcal{R}$ is defined with respect to the original metric $g_{ab}$. Here $(a,b)$ (i.e, the Latin alphabets in lower case), stand for all the spacetime coordinates $x\equiv (t,\bar{x})$.
We perform a general conformal transformation of the metric $g_{ab}$, as following,
\begin{equation}
g_{ab}(x)\, \to\, g'_{ab}(x)=\Omega^2(x)g_{ab}(x)\label{conformal_1}~,
\end{equation}
where $\O(x)$ is known as the conformal factor. Due to this, the determinant of the metric tensor, inverse metric, curvature scalar, will modify accordingly \cite{Birrell}: 
\beq
&\sqrt{g'}=[\O(x)]^{D} \, \sqrt{g}~;~~~~~~~~~g^{'ab}\,=\, \O(x)^{-2}\, g^{ab}~;\nonumber\\
& \mathcal{R'}=\, \O(x)^{-2}\mathcal{R}+(D-1)(D-4)\O^{-4}\, \O_{;a}\, \O_{;b}\,g^{ab}+2(D-1)\, \O^{-3}\O_{;ab}g^{ab}~,
\label{changes_1}
\eeq
where $D$ stands for the number of spacetime dimensions.
The conformal transformation of the spacetime metric also affects the scalar field present in this curved spacetime. Therefore, to obtain a canonical kinetic term for the scalar field, the field itself has to be redefined through a conformal transformation, as, 
\begin{equation}
\ph(x)\,\to\, \ph'(x)=\O^{(2-D)/2}(x)\, \ph(x)~,
\end{equation} 
along with the other changes as described in eqs. (\ref{changes_1}).
Now, if we impose that the action for a massless, non-minimally coupled scalar field, has to be invariant under the conformal transformation (as in eq. (\ref{conformal_1})), the non-minimal coupling strength can be uniquely fixed as,
\begin{equation}
\xi= \f{D-2}{4(D-1)}~,
\end{equation}
which vanishes for the special case $D=2$.

 In this present work, we consider that due to the conformal transformation, the original metric $g_{ab}$ transforms to that of  Minkowski spacetime as follows,  
\be
g_{ab}(x)\, \to \Omega^{-2}(x)g_{ab}=\eta_{ab}\equiv ~g'_{ab},
\label{conformal_min}
\ee
where $\eta_{ab}$ symbolises the metric corresponding to Minkowski spacetime.
 For a transformed metric (which is that of Minkowski spacetime), the modified scalar curvature becomes trivial and is, $\mathcal{R'}=0$ and the other quantities follow the changes as described in eq. (\ref{changes_1}) with $\Omega$ replaced by $\Omega^{-1}$. Similarly, the scalar get modified due to the conformal transformation as, 
 \begin{equation}
\ph'(x)=\O^{(D-2)/2}(x)\, \ph(x)~.
\label{scalar_modification}
\end{equation} 
Here the scalar field $\ph(x)$ corresponds to the original metric background (i.e, $g_{ab}$) and we denote $u_{\bar{k}}(x)$ as the positive frequency mode solutions of the corresponding field. The positive frequency mode solutions $u_{\bar{k}}(x)$ will also remain the same for the conformally transformed scalar field $\phi'(x)$ \cite{Birrell}. Consequently, the creation and the annihilation operators ($a^{\dagger}_{\bar{k}}$ and $a_{\bar{k}}$) become same for both the fields. Furthermore the vacuum state, defined by $a_{\bar{k}}\ket{0}=0$, also becomes a unique choice of vacuum state for both $\phi$ and $\phi'$ \cite{Birrell,Parker:1968,Parker:1969,Parker:1971}.
Therefore, for the field $\ph(x)$, one can define a suitable vacuum state which emerges naturally due to the conformal symmetry of the curved spacetime, and popularly known as the ``conformal vacuum'' (denoted as, $\ket{0}$). 
In this connection we mention that the Green function $D(x,x')$ of the massless scalar field, as evaluated  with respect to the conformal vacuum of the scalar field in the conformally flat spacetime $g_{ab}=\Omega^2\eta_{ab}$, is related to that ($D^{M}(x,x')$) for a massless scalar field located in its Minkowski vacuum (denoted as, $\ket{0_M}$) in flat Minkowski spacetime \cite{Birrell}. In general, this relation can be written as,
\be
D(x,x')=\O^{(2-D)/2}(x)\, D^{M}(x,x')\, \O^{(2-D)/2}(x')~.
\label{mod_green_1}
\ee
Later we shall use this relation to find the required Green's function for a spacetime which connected to the Minkowski spacetime through a conformal factor.

\subsection{Classic examples of the conformally flat spacetime}\label{examples}
Here we shall introduce the examples of curved spacetimes which can be expressed as conformally flat spacetime by going into a new coordinate system. We briefly explain particular examples which are relevant for our main analysis. 

{\underline{\it $(1+1)$-dimensional FLRW Universe}}: We initiate with $(1+1)$-dimensional spatially flat FLRW spacetime. The $(1+1)$-dimensional field theories provide significant predictions to formulate QFT in higher dimensions due to the analogy between the $(1+1)$ and $(3+1)$-dimensional QFT \cite{Birrell}. Upon using the conformal transformation as, $\e=\int_{0}^{t}\f{dt'}{a(t')}$ and $\Omega=a(\eta)$, the $(1+1)$-dimensional FLRW metric reduces to the conformally flat to Minkowski spacetime as, 
\be
ds^{2}_{(2)}=dt^2-a^2\,(t)[dx^2]\, \xrightarrow{\e=\int_{0}^{t}\f{dt'}{a(t')}} \, a^2(\eta)(d\eta^2-dx^2)~,
\label{frw_2d}
\ee
where $a(t)$ is the scale factor of the Universe and $\eta$ is known as the conformal time. It is evident from eq.      (\ref{scalar_modification}) and eq. (\ref{mod_green_1}) that for $D=2$, there is no contribution from the conformal factor, appearing in the transformation of the scalar field and as well as the Green function. Therefore the scalar field remain same as the original metric background and the Green function $D(x,x')$ turns out to be same as the Green function $D^{M}(x,x')$, in $(1+1)$-dimensional spacetime. This type of metric has been considered earlier in several situations to study the particle production phenomenology (e.g. see \cite{Birrell}; also see section $3.1$ of \cite{Singh:2013pxf} for a recent study).

{\underline{\it $(3+1)$-dimensional FLRW Universe}}: Another example of the curved spacetime with conformal symmetry is the $(3+1)$-dimensional spatially flat FLRW metric. After performing the conformal transformation as in the previous case, the $(3+1)$-dimensional FLRW metric reads as :
\begin{equation}
ds^{2}=dt^2-a^2\,(t)[dx^2+dy^2+dz^2]\,  \xrightarrow{\e=\int_{0}^{t}\f{dt'}{a(t')}} \,
a^2(\eta)(d\eta^2-dx^2-dy^2-dz^2)~.
\label{frw_1}
\end{equation}
For $D=4$, the scalar field transforms under conformal transformation as, $\ph'(x)=\O(x)\, \ph(x)$ and the modification in the Green function becomes, 
\be
D(x,x')=\O^{-1}(x)\, D^{M}(x,x')\, \O^{-1}(x')~.
\label{mod_green_4d}
\ee

{\underline{\it $(1+1)$-dimensional Schwarzschild BH}}: We urge to cite $(1+1)$-dimensional Schwarzschild spacetime, as another example of curved spacetime which can be reduced to conformally flat to Minkowski spacetime. The Schwarzschild metric can be written in its usual form as, 
\begin{eqnarray}
ds^2_{S} = \left(1 - \frac{r_s}{r} \right) \: dt^2  -  \left( 1 - \frac{r_s}{r} \right)^{-1}\: dr^2~. 
\label{sch_1}
\end{eqnarray}
Here, $r_s=2M$
 is the Schwarzschild radius and $M$ is the mass of the BH. Performing the coordinate transformation as, $r_{*}=r+r_{s}\ln\left(\frac{r}{r_{s}}-1\right)$, the above metric reduces to the conformally flat metric to Minkowski spacetime with the coordinates $(t,r_{*})$.  Further performing another coordinate transformation as,
\begin{eqnarray}
\bar{u}=-\,4M\: e^{-\frac{u}{4M}}, \:\:\: \bar{v}=\,4M\: e^{\frac{v}{4M}}\label{KS_1}
\end{eqnarray}
the metric reduces to the following form,
\begin{eqnarray}
ds^{2}_{K}=\frac{r_{s}}{r} \, e^{-\frac{r}{r_{s}}}\,d\bar{u}\,d\bar{v}~.
\label{sch_kruskal_1}
\end{eqnarray}
Here, $t+r_*=v$, $t-r_*=u$ are the null coordinates. In the two-dimensional Schwarzschild spacetime, the scalar field will remain unchanged under the conformal transformation and the Green function $D(x,x')$ comes out to be same as the Green function $D^{M}(x,x')$. Same metric as in eq.(\ref{sch_kruskal_1}), has also been considered in \cite{Birrell} to study the phenomena of particle production in different vacuum states.

Note that here we shall discuss $(1+1)$-dimensional situation for both FLRW and BH spacetimes. These models have been considered purposefully. There are a couple of reasons for this. First of all, $(1+1)$-dimensional cases are simple and analytically solvable. Secondly, as far as particle production and thermal behaviour of field modes are concerned (like temperature), this study almost complementary to the system in actual higher dimensions. Therefore, the outcomes in $(1+1)$-dimensional spacetimes also provide a close understanding of the results of higher-dimensional spacetimes.  
For example, it is well known that in the derivation of Hawking radiation, it is very hard to define the wave functions corresponding to a quantum matter field in the spacetime geometry of a collapsing body. Therefore, to define the mode functions of the field, a two-dimensional spherically symmetric spacetime metric is considered in order to depict a shrinking ball of matter \cite{Birrell}. It is stated in \cite{Birrell}, that working with two-dimensional spacetime is also beneficial in terms of the renormalizability of the theory at all spacetime points. 
Similarly, the phenomena of particle production and the origin of thermal bath in a particular choice of vacuum state can be realised in a simplified manner in the background of $(1+1)$-dimensional cosmological spacetimes such as FLRW metric as in eq.(\ref{frw_2d}) \cite{Birrell}. So it is obvious that $(1+1)$-dimensional studies require special attention, and indeed this has been the case from the very beginning of the quantum field theory in curved spacetime. Two-dimensional Schwarzschild case has been extensively studied starting from particle production to Hawking radiation (e.g. see \cite{Balbinot:1999vg}).
In this regard, it may be mentioned that the Schwarzschild metric (\ref{sch_1}), can be obtained as a solution of dimensionally reduced Dilaton theory from $(3+1)$ to $(1+1)$ dimensions \cite{Grumiller:2002nm}. Moreover, there is a physical justification of considering a metric like (\ref{sch_1}), which is as follows. It is well known that the theory of fields in the background of $(3+1)$-dimensional spacetime, reduces to conformal theory on an effectively $(1+1)$-dimensional background of the form identical to (\ref{sch_1}) \cite{Carlip:1998wz, Robinson:2005pd, Iso:2006wa, Majhi:2011yi}, in the near horizon region. Incidentally, this effective metric is sufficient to study the emission of the particle from the horizon \cite{Robinson:2005pd, Iso:2006wa}, since the emission of the particle, in general, is a near horizon phenomenon. These utilities and simplicity of $(1+1)$-dimensional theories, have led us to investigate the behaviour of the produced particles in the thermal bath, by considering $(1+1)$-dimensional spacetime as our model background. 
Of course, investigations in $(3+1)$-dimensional spacetime will lead to more detailed information about the system which we leave for future. As a beginning, here we concentrate on some simple situations. In this regard, it may be worth to point out that there are other types of BH solutions existing in $(1+1)$ dimensions \cite{Lemos:1994fn, Ref2, Ref3} whose thermal behaviour can also be investigated similar to our approach.

\subsection{Relevant observers and particle detection}
So far we have discussed that the conformal vacuum is a natural choice of vacuum state in conformally flat curved spacetime.   However, one needs to be more careful before jumping to any conclusion as, it is nowhere mentioned in the above discussion, that what is the state of motion of the measuring device which is employed to detect the particle states. Here we shall discuss the same for the above three spacetimes. 

\subsubsection{de Sitter universe}
In literature \cite{Birrell,Parker:1968,Parker:1969,Parker:1971,Gibbons:1977,MartinMartinez:2012sg} a ``privileged class of observers'' have been defined in the FLRW spacetime, known as comoving observer. Comoving observers see the homogenous and isotropic expansion of the Universe with respect to their proper reference frame, where the comoving coordinates are represented by ($t,\, \bar{x}$).
The particle production in the conformal vacuum can be established by implementing the idea of the monopole interaction of a point-like Unruh-DeWitt detector and a scalar field, located in its Minkowski vacuum \cite{Birrell,Unruh:1973,DeWitt:1975}. 
Due to this monopole interaction, the power spectrum/response function of the comoving detector turns out to be related to the two-point correlation function of the scalar field i.e, the positive frequency Wightman function $G^{+}(x,x')$, evaluated with respect to the conformal vacuum of the field \cite{Birrell,Davies:1974,Bunch:1978}. 
For the dS Universe, the scale factor of the Universe is $a(t)=e^{t/\a_{d}}$, where $\a_{d}=H^{-1}$, is known as the inverse Hubble expansion rate. The conformal and the comoving time is related as : $\eta=-\a_{d}\, e^{-t/\a_{d}}$ and $\O(x)=a(\eta)=\a_{d}/\eta$.

In $(1+1)$ dimension the positive frequency Wightman function corresponding to the massless scalar field in conformally flat spacetime becomes,
  \begin{eqnarray}
G^{+}_{(2)}(\eta,x;\eta',x')=-\frac{1}{4\pi}\log\left[(\eta-\eta')^{2}-(\bar{x}-\bar{x}')^{2}\right]~.
\label{green_2d_frw}
\end{eqnarray}
Similarly, form eq. (\ref{mod_green_4d}), the positive frequency Wightman function for $(3+1)$-dimensional FLRW spacetime turns out to be, 
\be
G^{+}(x,x')=\f{-\eta \eta' }{4\pi^2\a^2_{d}\bigg[\bigg(\Delta\eta-i\, \epsilon\bigg)^2-|\Delta \bar{x}|^2\bigg]}~,
\label{mod_green_ds}
\ee
where we use the known expression for the Green function with respect to the Minkowski vacuum of a massless scalar filed in Minkowski flat spacetime, which in $(3+1)$ dimension varies as, $\sim -\f{1}{4\pi^2[(\Delta\eta-i\, \epsilon)^2-|\Delta \bar{x}|^2]}$ and for $(1+1)$ dimension is same as that of the eq. (\ref{green_2d_frw}). 

The response function has to be evaluated with respect to the proper frame of the comoving detector. Therefore, we consider, $|\Delta \bar{x}|=\bar{x}-\bar{x}'={\rm constant}=0$, and in the background of FLRW spacetime, the proper time ($\ta$) of the comoving detector coincides with the coordinate time ($t$).  
Therefore, one finds the response function per unit time of the comoving detector in both the $(1+1)$ and $(3+1)$ dimensions varies as \cite{Birrell},
\be
\mathcal{F}(\Delta E)\sim \,\f{ \mathfrak{g}}{e^{2\pi\a_{d} \Delta E}-1}~,
\label{response_2}
\ee
where $\mathfrak{g}$ is a pre-factor, which in $(1+1)$ dimensions becomes $\f{1}{\Delta E}$ [see chapter 4 of \cite{Takagi:1986kn} for detail discussions] 
 and in $(3+1)$ it is given by $\f{\Delta E}{2\pi}$, with $\Delta E$ is the difference in the initial and the final energy eigenvalue of the comoving detector.
Therefore a comoving detector in the dS Universe detects thermal spectrum (i.e., particle excitations) in the conformal vacuum of the scalar field in conformally flat spacetime.
The temperature of the bath, $T$, is related to the curved space parameters as, $T=\f{1}{2\pi\a_{d}}$, where the Boltzmann constant ($k_B$), Plank's constant ($\hbar$) and velocity of light ($c$) are set to be $1$. Additionally, we mention that the above observation can also be realised by calculating either the vacuum expectation value of the number operator or the vacuum expectation value of the time-time component (i.e., $T^{tt}$) of the energy-momentum tensor of the corresponding quantum field, with respect to the conformal vacuum. For completeness, we evaluate the renormalised expectation value of the $T_{t}\,^{t}$ component of the stress-energy tensor with respect to the conformal vacuum in Appendix \ref{app1}. Indeed, we obtain a non-zero finite energy density corresponding to this temperature, which signifies the particle production in the conformal vacuum from the point of view of a comoving observer.

\subsubsection{$(1+1)$-dimensional Schwarzschild spacetime} \label{description_sch}
We consider a massless quantum scalar field in the background of $(1+1)$-dimensional Schwarzschild spacetime where, as similar to the FLRW spacetime, some vacuum states of the quantum field emerge naturally. These vacuum states are known as, Boulware vacuum ($\ket{0}_s$), Kruskal vacuum ($\ket{0}_K$) and Unruh vacuum ($\ket{0}_{U}$) \cite{Birrell}. For a detailed discussion of the emergence of these vacuum states and its significance, we refer our readers to \cite{Birrell}. These vacuum states act as the conformal vacuum in the present discussion in their respective coordinate systems; i.e. $\ket{0}_s$ in $(u,v)$, $\ket{0}_K$ in $(\bar{u},\bar{v})$ and $\ket{0}_U$ in $(\bar{u},v)$. Proceeding similarly as in the last section, one can study further the behaviour of these vacuum states with respect to a particular observer by using the technique of detector-field interaction. 

Following the discussion for Schwarzschild spacetime as in Section \ref{examples}, the positive frequency Wightman functions, evaluated with respect to the Boulware, Kruskal and Unruh vacuum states can be written as \cite{Birrell}, 
\be
G^{+}_{s}\,(x,x')=-\,\f{1}{4\pi}{\rm ln}[(\Delta u-i\epsilon)(\Delta v-i\epsilon)]~,\label{green_sch}
\ee
\be
G^{+}_{K}\,(x,x')=-\,\f{1}{4\pi}{\rm ln}[(\Delta \bar{u}-i\epsilon)(\Delta \bar{v}-i\epsilon)]~,\label{green_kruskal}
\ee
\be
G^{+}_{U}\,(x,x')=-\,\f{1}{4\pi}{\rm ln}[(\Delta \bar{u}-i\epsilon)(\Delta v-i\epsilon)]~.
\label{green_unruh}
\ee
Consider the trajectory of a particle detector to be constant $r$, say  $r=R$. The proper time in this frame is $d\ta=(1-\f{2\, M}{R})^{1/2}\, dt$. The response function of the detector, while the quantum scalar field is located at $\ket{0}_s$, reduces to zero, i.e, {\it the detector records no particle excitations in the Boulware vacuum state}. 
The response function per unit time of the detector in case of the scalar field at Unruh vacuum and Kruskal vacuum turns out to be thermal by nature, as similar to eq. (\ref{response_2}). The temperature of the thermal bath becomes, $T_U=[64\pi^2\, M^2 (1-\f{2M}{R})]^{-1/2}$ and it can be shown that $T_U$ is related with the surface gravity ($\kappa$) of the Schwarzschild BH, since, $\kappa=1/(4M)$. In order to realise the thermal spectrum in terms of the non-zero value of the energy density in the corresponding vacuum states, we evaluate the expectation value of $T_{t}\,^{t}$ component of the stress-energy tensor with respect to the Unruh/Kruskal vacuum in Appendix \ref{app3}.

\section{Setup: Definitions of momentum and force-force correlation}\label{methodology}
In the upcoming sections, we examine the behaviour of the produced quanta in the particular vacuum state, from the point of view of the privileged class of observer.
 We follow the same procedure as adopted in \cite{Bibhas_fdt}, which is reminiscent to an earlier work \cite{Ford:1982ct}. Firstly, we define the random force, exerted by these produced quanta on the test particles, as measured by a particular class of observer. This observer is indeed capable to detect the scalar field quanta in the conformal vacuum. Subsequently, we calculate the correlation function of the fluctuations of the random force with respect to the chosen vacuum state. The force-force correlation function carries the information regarding the guiding principle, as followed by the produced particles in the corresponding vacuum state.

\subsection{Random force} 
It is well known that the physical velocity of a particle, as measured by a comoving observer in FLRW spacetime, can be expressed as $v^{i}_{\rm{phys}}=a(t)\, \frac{dx^{i}}{dt}$ \cite{Baumann}. We generalise this concept in order to define the momentum of a quantum scalar field in the FLRW and BH spacetime. 
Therefore in FLRW spacetime, we define the conserved three momenta of these produced scalar quanta, in terms of the stress-energy tensor of the corresponding field, measured by the comoving observer, as following: 
\be
P^{\a}=L^na(\tau)\int T^{t\a} (\tau,\bar{x}) \d(\bar{x}-\bar{x}_D) d^nx =L^na(\tau)\, T^{t\a}(\ta)~,\label{momentum_comoving_1}
\ee
where $\a$ symbolises the spatial coordinates and $n$ denotes the number of spatial dimensions. Here $t=\tau$ is the proper time of the detector. We mention that all the Greek letters except $\eta$, stand for the spatial coordinates of the corresponding spacetime. Here we insert the Dirac delta function in order to incorporate the information that the above quantity is measured with respect to the coordinates $x_D=(t= \tau, \ \bar{x}_D=0)$ of the detectors. Hence the momentum becomes the function of the proper time ($\ta$) of the detector. In the above equation, $L$ is some length parameter of the system which has been introduced in order to get the dimensions right. This can be the size of the detector for our present setup. Later on, we shall calculate all the quantities in units of volume corresponding to this length parameter and loosely call them as momentum, force, etc. So we define our ``momentum'' (per unit volume) as
\be
p^{\a}=a(\tau)\, T^{t\a}(\ta)~.
\label{momentum_comoving}
\ee
For $(1+1)$-dimensional case, it would be per unit length.
Similarly, one can define the momentum associated with the emitted scalar quanta from the Kruskal, Unruh vacuum as measured by the detector in $(1+1)$-dimensional Schwarzschild spacetime as
\be
p^r=\int T^{tr} \d({r-r_D}) dr =\, T^{tr}(\ta)~,
\label{momentum_sch}
\ee
where $r$ symbolises the radial coordinate corresponding to the Schwarzschild metric.

The force as exerted by the emitted quanta of the scalar field and measured by the comoving observer is defined as,\footnote{This relation as of now is purely phenomenological. 
This definition is applicable to all the curved backgrounds of our interest in this present work, irrespective of the dimensionality of the spacetime. 
It may be possible to give a microscopic derivation of the same by modelling the interaction between the test particle and the detector more robustly, however, this point shall be addressed in a future work.}
\be
F^{\a}\equiv\frac{dp^\alpha}{d\tau} = \f{d}{d\ta}(a(\ta)T^{t\a}(\ta))~,
\label{force_comoving}
\ee
and similarly for the BH case.
In order to find the correlation function of the random force, one needs to define the random part of the force. This is defined as
\be
R^{\a}(\ta)=\,F^{\a}(\ta)-\langle F^{\a}(\ta)\rangle
\label{fluctuation_1}
\ee
where, $\langle F^{\a}(\ta)\rangle$ symbolises the vacuum average of the force. Subsequently, the desired correlation function is quantified as: $\bra{0}R^{\a}(\ta)R^{\a}(\ta')\ket{0}$.

 \subsection{Mathematical steps to obtain the correlation function}\label{mathematical_analysis}
\begin{itemize}
\item
For FLRW, the correlation function $\bra{0}R^{\a}(\ta)R^{\a}(\ta')\ket{0}$ can be written in terms of the correlation function of the corresponding stress-energy tensors as $\sim \langle{0}\vert T^{t\a}(t,\bar{x})T^{t\a}(t',\bar{x}')\vert{0}\rangle$. 
We find it more convenient to work with $\bra{0}T^{\e \a}(\e,\bar{x})T^{\e \a}(\e',\bar{x}')\ket{0}$.
For $(3+1)$-dimensional spacetime, we transform $T^{t\a}$ to $T^{\eta \a}$ i.e, in terms of the coordinates of the conformally flat spacetime by tensorial transformation as follows,
\begin{eqnarray}
T^{t\a}(t,\bar{x})&&= \frac{\partial t}{\partial \eta}\frac{\partial x^{\a}}{\partial x^{\beta}}\: T^{\eta\beta}(\eta,\bar{x})\nonumber
\\
&&= \frac{\partial t}{\partial \eta}\delta^{\a}_{\beta}\: T^{\eta\beta}(\eta,\bar{x})= \frac{\partial t}{\partial \eta}T^{\eta\a}(\eta,\bar{x})~.
\end{eqnarray}
\item
Upon using the explicit form of the stress-tensor in the corresponding background metric and implementing the Wick's theorem, the correlation function $\bra{0}T^{\e \a}(\e,\bar{x})T^{\e \a}(\e',\bar{x}')\ket{0}$, can be further reduced to the positive frequency Wightman function of the scalar field, evaluated with respect to the conformal vacuum in the conformally flat spacetime.
\item
We exploit the conformal symmetry of this curved background and obtain the positive frequency Wightman function of the massless scalar field in the conformally flat background by using the relation in eq. (\ref{mod_green_1}). 
\item
The correlator comes out to be the function of the conformal coordinates i.e, for example, ($\eta, \bar{x}$) for FLRW spacetime. Therefore, finally one needs to transform the correlator in terms of the proper coordinates of the comoving observer i.e, $(t,\bar{x})$ (for FLRW), by performing the coordinate transformation and then obtain the desire correlation function $\bra{0}R^{\a}(\ta)R^{\a}(\ta')\ket{0}$. 
\end{itemize}
Similar steps also follows for the Schwarzschild case.
In the upcoming section we implement the described methodology in order to find the correlation function of the fluctuation of the random force in the context of the $(1+1)$ \& $(3+1)$-dimensional FLRW spacetime and $(1+1)$-dimensional Schwarzschild spacetime and also explore the nature of the produced particles in the corresponding vacuum state. 
\section{Correlators corresponding to the random force}
Having the prescribed steps to evaluate the correlators, we are now in a position to obtain them. In this section, we calculate them explicitly for the cases under study in this paper.
\subsection{de Sitter Universe}\label{FD_cosmology}
\subsubsection{(1+1) dimensions}
In $(1+1)$ dimensions the only possible momentum component is $T^{tx}$, which signifies the momentum along the x-direction and corresponding $F^{x}(\ta)$, is the force along the x-direction.
One can show that the renormalised vacuum average of the force i.e,  $\langle F^{x}(\ta)\rangle$, turns out to be zero, as $\langle T^{tx}(\tau)\rangle =0$. We refer our readers to the Appendix \ref{app2}, for a detailed analysis. Therefore the fluctuation of the random force reduces to, $R^{\a}(\ta)=\,F^{\a}(\ta)$.  
Thus, the correlation function of the fluctuation of the random force in $(1+1)$-dimensional FLRW spacetime becomes,
\begin{eqnarray}
\langle 0|R^{x}(\tau)R^{x}(\tau ')|0\rangle=\frac{d}{d\tau}\frac{d}{d\tau '}\left[a(t)a(t')\langle{0}\vert T^{tx}(t,\bar{x})T^{tx}(t',\bar{x}')\vert{0}\rangle\right]~.
\label{correlation_flw_1}
\end{eqnarray}
From this point, we follow all the  steps as described under the subsection  \ref{mathematical_analysis}. Thus in $(1+1)$ dimension, we obtain, $T^{tx}(t,x)=-\f{\a_{d}}{\e}\, T^{\e x}(\e,x)=\frac{1}{a^3(\e)}T_{\e x} (\e,x)$ and subsequently, eq. (\ref{correlation_flw_1}) reduces to the following form:
\begin{align}
\langle 0|R^{x}(\tau)R^{x}(\tau ')|0\rangle=\frac{d}{d\tau}\frac{d}{d\tau '}\bigg[\f{1}{a^2(\e)a^2(\e')}&\bigg\{\partial_{\e}\,\partial_{\e'}\,G^{+}_{(2)}(\e,x;\e',x')\nonumber\\
&\qquad \times \partial_{x}\,\partial_{x'}\,G^{+}_{(2)}(\e,x;\e',x')\bigg\}\bigg]~.
\label{correlation_flw_2}
\end{align}
To reach the final expression of the above, see Appendix \ref{explicit derivation_em}.
Finally we use the proper coordinates of the comoving observer i.e, $|\Delta \bar{x}|=\bar{x}-\bar{x}'={\rm constant}=0$ and $t=\ta$, in the above correlation function. 
Upon using the eq. (\ref{green_2d_frw}), and the explicit form of $a[\e(t)]$, the above equation becomes
\begin{eqnarray}
&&\langle 0|R^{x}(\tau)R^{x}(\tau ')|0\rangle =
\frac{d}{d\tau}\frac{d}{d\tau '}\bigg[\frac{e^{-\frac{2t}{\alpha_{d}}}\:e^{-\frac{2t'}{\alpha_{d}}}}{4\:\pi^{2}}\times\frac{1}{(\eta-\eta ')^{4}}\bigg]
\nonumber
\\
&&=\frac{1}{2^{6}\:\alpha^{4}_{d}\:\pi^{2}}\bigg[\frac{d}{d\tau}\frac{d}{d\tau '}\bigg\{\frac{1}{\sinh^{4}{\frac{\Delta \tau}{2\:\alpha_{d}}}}\bigg\}\bigg]
 =-\frac{1}{2^{6}\:\alpha^{6}_{d}\:\pi^{2}}\times\frac{\left[5+4\sinh^{2}{\frac{\Delta\tau}{2\:\alpha_{d}}}\right]}{\sinh^{6}{\frac{\Delta\tau}{2\:\alpha_{d}}}}~,
 \label{correlation_flw_4}
\end{eqnarray}
where, $\Delta\tau=\tau-\tau'$.

At this stage, we close the discussions on $(1 + 1)$ spacetime and proceed to the analysis of the same in the context of $(3+1)$-dimensional FLRW spacetime. We analyse the behaviour of the correlation function (i.e, eq. (\ref{correlation_flw_4})) in the later parts of Section \ref{fd_property}. In Appendix \ref{Schwinger}, we explicitly calculate the correlation function of the stress-energy tensor, using the Schwinger function. The result exactly matches with the second last expression of the eq. (\ref{correlation_flw_4}).
The exact equality of the two results which were obtained via two different methods, strengthen our analysis of force-force correlation in $(1+1)$-dimensional de Sitter FLRW spacetime.
Also it must be noted that the Schwinger function cannot be implemented in case of the $(3+1)$-dimensional FLRW spacetime. This is because, all the results involving the same are only applicable for a specific form of $T_{ab} \sim \p_a \phi \p_b \phi$. In $(3+1)$ dimensions, due to the non zero non-minimal coupling strength $\xi$, there would be additional terms along with $ \p_{a}\phi\p_{b}\phi$, in the explicit form of $T_{ab}$ (we refer to eq. (\ref{gen_em_1})). Therefore one fails to implement the Schwinger function in $(3+1)$-dimensional FLRW case. 
\subsubsection{(3+1) dimensions}
In $(3+1)$-dimensional FLRW there are three component of the force along the $(x,y,z)$ directions and accordingly one needs to compute the momentum component such as $(T^{tx},T^{ty},T^{tz})$. For all three directions, following the similar steps as above, one finds that the value of random force correlator is same :
\begin{align}
\langle 0|R^{\alpha}(\tau)R^{\alpha}(\tau ')|0\rangle &=\frac{d}{d\tau}\frac{d}{d\tau '}\bigg[\f{1}{a^2(\eta)a^2(\eta')}\,\f{\eta^2\eta^{'2}}{6\alpha_{d}^{4}\pi^4\,(\eta-\eta')^8}\bigg]\label{deriv_stress}\\
&=-\frac{1}{2^{7}\:\alpha_{d}^{10}\:(6\pi^{4})}\times\frac{\left[9+8\sinh^{2}{\frac{\Delta\tau}{2\alpha_{d}}}\right]}{\sinh^{10}{\frac{\Delta\tau}{2\alpha_{d}}}}~.
\label{correlation_flw_5}
\end{align}
This is because the FLRW universe is homogeneous and the isotropic, and hence any quantity in all directions should be same. For details of the calculation, we refer to Appendix \ref{explicit derivation_em}.
The quantity $\Delta\ta$ symbolises the proper time, like that of the $(1+1)$-dimensional case.
As stated earlier, further analysis of the behaviour of the correlation function, i.e., eq. (\ref{correlation_flw_5}), will be discussed in Section \ref{fd_property}. 
\subsection{($1+1$) dimensional Schwarzschild BH}\label{FD_blackhole}
We observed that the static observer in Schwarzschild coordinates would not see any particles in the Boulware vacuum while in the other two vacuums the same observer detects particles. 
Therefore, here we focus on finding the correlation function of the fluctuations of the random force by the produced particles in the Kruskal, Unruh vacuum, as observed by this privileged class of observer in the Schwarzschild spacetime. 

We write the Schwarzschild metric eq. (\ref{sch_1}), in terms of the Unruh coordinates $(\bar{u},v)$:
\begin{eqnarray}
ds^{2}_{U}=\left(1-\frac{r_{s}}{r}\right)\left(-\frac{4M}{\bar{u}}\right)\:\:d\bar{u}\:\:dv~,
\label{sch_unruh}
\end{eqnarray} 
whereas the same in Kruskal coordinates, $(\bar{u},\bar{v})$ takes the form eq. (\ref{sch_kruskal_1}).
Following the discussions in methodology section, the $T^{tr}(t,r)$ component of the energy momentum tensor 
in the Schwarzschild spacetime can be written in terms of the coordinates $(\bar{u},\bar{v})$ and $(\bar{u},v)$. Therefore $T^{tr}(t,r)$ component reduces to,
\begin{eqnarray}
T^{tr}_{K}&&=4M^{2}\left(1-\frac{r_{s}}{r}\right)\left[\frac{T^{\bar{v}\bar{v}}}{\bar{v}^{2}}-\frac{T^{\bar{u}\bar{u}}}{\bar{u}^{2}}\right]\nonumber\\
&&=4r^{2}\left(1-\frac{r_{s}}{r}\right)\,e^{\frac{2\,r}{r_{s}}}\,\bigg[\frac{T_{\bar{u}\bar{u}}}{\bar{v}^{2}}-\frac{T_{\bar{v}\bar{v}}}{\bar{u}^{2}}\bigg]~;
\label{stress_kruskal}
\end{eqnarray}
and 
\begin{eqnarray}
T^{tr}_{U}&&=\frac{1}{4}\left(1-\frac{r_{s}}{r}\right)\left[T^{vv}-\frac{16\:M^{2}}{\bar{u}^{2}}\:T^{\bar{u}\bar{u}}\right]\nonumber\\
&&=\left(1-\frac{r_{s}}{r}\right)^{-1}\bigg[\left(\frac{\bar{u}}{4M}\right)^{2}T_{\bar{u}\bar{u}}-T_{vv}\bigg]~,
\label{stress_unruh}
\end{eqnarray}
where, $T^{tr}_{K},\,T^{tr}_{U}$ are the stress-energy tensor of the scalar field corresponding to the Kruskal and Unruh vacuum. The above two equations are containing two parts: $T_{\bar{u}\bar{u}}$ corresponds to the outgoing radiation flux while $T_{\bar{v}\bar{v}},\,T_{vv}$ is the ingoing radiation flux of the produced particles from the respective vacuum states. Since the outgoing modes can only be perceived by the observer in the Schwarzschild spacetime, we concentrate only to the outgoing modes of the eqs. (\ref{stress_kruskal}) and (\ref{stress_unruh}). 

In case of the Kruskal vacuum the required component of stress-tensor for the outgoing modes is,
\begin{eqnarray}
\f{4r^{2}}{\bar{v}^{2}}\left(1-\frac{r_{s}}{r}\right)e^{\frac{2r}{r_{s}}}\,T_{\bar{u}\bar{u}}
\end{eqnarray}
and the corresponding stress tensor correlator becomes,
\begin{align}
_{K}\langle 0|T^{tr}_{K}(t,r)T^{tr}_{K}(t',r')|0\rangle_{K}
 &=16r^{2}r^{'2}\left(1-\frac{r_{s}}{r}\right)\left(1-\frac{r_{s}}{r'}\right)e^{\frac{2r}{r_{s}}}\,e^{\frac{2r'}{r_{s}}}
 \nonumber
\\
&\qquad \times\frac{1}{\bar{v}^{2}\bar{v}'^{2}}
\langle 0|T_{\bar{u}\bar{u}}(\bar{u},\bar{v})T_{\bar{u}\bar{u}}(\bar{u}'\bar{v}')|0\rangle \nonumber
\\
&=\frac{1}{(8)^5\: \pi^{2}M^4}\left(1-\frac{r_{s}}{R}\right)^{-2}\times\frac{1}{\sinh^{4}\left(\frac{\Delta\tau}{8M\sqrt{1-\frac{r_s}{R}}}\right)}~.
\label{correlation_sch_k}
\end{align}
To compute the above equation, we proceed similarly as described in Section \ref{description_sch} and use the positive frequency Wightman function as in eq. (\ref{green_kruskal}). The trajectory of the observer in the Schwarzschild spacetime, is set to be $r=R$ (constant curvature) and $\tau$ symbolises the proper time of the detector in the Schwarzschild spacetime. $\Delta \ta$ symbolises the same as that of the earlier sections. 
Therefore the correlation function for the fluctuations of the random force, $_{K}\langle 0|R(\tau)R(\tau ')|0\rangle_{K} $, as measured by the static observer in the Schwarzschild spacetime becomes,
\begin{eqnarray}
_{K}\langle 0|R^{r}(\tau)R^{r}(\tau ')|0\rangle_{K} &&=\frac{d}{d\tau}\frac{d}{d\tau '}\left[\frac{\left(1-\frac{r_{s}}{R}\right)^{-2}}{(8)^5\: \pi^{2}M^4}\times\frac{1}{\sinh^{4}\left(\frac{\Delta\tau}{8M\sqrt{1-\frac{2M}{R}}}\right)}\right]\nonumber
\\
&&=-\frac{\left(1-\frac{2M}{R}\right)^{-3}}{2\: \pi^{2}\: (8\: M)^{6}} \times\frac{\left[5+4\sinh^{2}{\left(\frac{\Delta\tau}{8M\sqrt{1-\frac{2M}{R}}}\right)}\right]}{\sinh^{6}{\left(\frac{\Delta\tau}{8M\sqrt{1-\frac{2M}{R}}}\right)}}~.
\label{correlation_sch_k1}
\end{eqnarray}

We perform the same analysis in the context of the particle production in Unruh vacuum and obtain the stress tensor correlator for the outgoing modes as exactly same as obtained in case of the Kruskal vacuum analysis, depicted in eq. (\ref{correlation_sch_k1}). 
\section{Nature of the correlation functions}\label{fd_property}
In this section, we focus on the examination of the nature of the obtained correlation functions of the fluctuations of the random force. Being motivated from the earlier discussions related to the moving mirrors \cite{Sriramkumar} and the Unruh radiation \cite{Bibhas_fdt}, we urge to verify that whether the produced particles in the curved backgrounds of our interest as discussed in the present work, follow the same FDT. 
Before proceeding to the analysis, we provide a brief description of the FDT, and for an elaborate discussion, we refer our readers to \cite{Kubo}. 

For any general operator $\mathcal{\hat{O}}$ the correlation function can be written as,
\be
K_{\mathcal{O}}(t)=\langle \mathcal{\hat{O}}(t_0)\,\mathcal{\hat{O}}(t_0+t) \rangle~,
\label{corre_fdt}
\ee
provided the explicit evaluation of the right hand side is time translational invariant. 
With the above definition, one can also define the symmetric and the antisymmetric correlation function of the operator $\mathcal{\hat{O}}$ as following \cite{Reif},
\beq
K_{\mathcal{O}}^{+}(t) &&\equiv \f{1}{2}\bigg[\langle \mathcal{\hat{O}}(t_0)\, \mathcal{\hat{O}}(t_0+t) \rangle+\langle \mathcal{\hat{O}}(t_0+t)\, \mathcal{\hat{O}}(t_0) \rangle\bigg]\nonumber\\
&&=\f{1}{2}\bigg[K_{\mathcal{O}}(t)+K_{\mathcal{O}}(-t)\bigg]~;
\label{corre_fdt_1}
\eeq
and
\beq
K_{\mathcal{O}}^{-}(t) &&\equiv \f{1}{2}\bigg[\langle \mathcal{\hat{O}}(t_0)\, \mathcal{\hat{O}}(t_0+t) \rangle-\langle \mathcal{\hat{O}}(t_0+t)\, \mathcal{\hat{O}}(t_0) \rangle\bigg]\nonumber\\
&&=\f{1}{2}\bigg[K_{\mathcal{O}}(t)-K_{\mathcal{O}}(-t)\bigg]~.
\label{corre_fdt_2}
\eeq
With these above definitions, the FDT can be described as: {\it if the Fourier transformation of $K_{\mathcal{O}}^{+}(t),\,K_{\mathcal{O}}^{-}(t)$ i.e, the symmetric and the antisymmetric correlators in the frequency space, satisfy the following relation, 
\be
\tilde{K}^{+}_{\mathcal{O}}(\o)\,={\rm coth}\bigg(\f{\b\o}{2}\bigg)\,\tilde{K}^{-}_{\mathcal{O}}(\o)\label{fdt_1}
\ee
the random function $\mathcal{\hat{O}}$, obeys the FDT}. In the above expression $\o$ symbolises the frequency of the corresponding operator $\mathcal{\hat{O}}$ in the phase space. 
$\tilde{K}^{+}_{\mathcal{O}}(\o),\,\tilde{K}^{-}_{\mathcal{O}}(\o)$ symbolise the symmetric, antisymmetric correlators in the frequency domain. Here, $\b=\f{1}{T}$ and $T$ signifies the temperature of the thermal bath of the produced particles.

Now, we examine the Fourier transformed version of the correlation functions of the random forces in our cases and symbolise $K(\ta)=\langle R^{\s}(0)\,R^{\s}(\ta)\rangle$ in the time domain. Here the initial time is set to zero for the sake of simplicity and this is admissible as the correlators are time translational invariant.  
The Fourier transform of this quantity is
\begin{eqnarray}
K(\omega)=\int_{-\infty}^{\infty}d\tau\: e^{i\omega\tau}\langle R^{\s}(0)R^{\s}(\tau)\rangle~.
\end{eqnarray}
It is evident from eqs. (\ref{correlation_flw_4}), (\ref{correlation_flw_5}) and (\ref{correlation_sch_k1}) that the obtained correlation functions are proportional to the inverse of the even powers of ${\rm sinh}$ function. 
\subsection{de Sitter Universe}\label{desitter_nature}
\subsubsection{$(1+1)$ dimensions}
We begin with the Fourier transformation of the correlation function eq. (\ref{correlation_flw_4}) corresponding to the $(1+1)$-dimensional FLRW de-Sitter spacetime as following,
\begin{align}
K(-\omega)=-\f{1}{2^6\alpha_{d}^{6}\pi^2}&\bigg[\underbrace{\int_{-\infty}^{\infty}d\tau\: e^{-i\omega\tau}\times\frac{5}{\sinh^{6}\bigg(\f{\ta}{2\alpha_d}\bigg)}}_{I_{1}} \  +\  \underbrace{\int_{-\infty}^{\infty}d\tau\: e^{-i\omega\tau}\times\frac{4}{\sinh^{4}{\bigg(\f{\ta}{2\alpha_d}\bigg)}}}_{I_{2}}\bigg]~.
\label{fdt_de_1}
\end{align}
The above integral is of the form as $I=\int_{-\infty}^{\infty}\frac{e^{-i\rho x}\,dx}{\sinh^{2n}{(x-i\epsilon)}}$, and widely used in the literature \cite{Gradsteyn,Paddy}. The integral can be reduced to the compact form as follows,
\begin{eqnarray}
I=\int_{-\infty}^{\infty}\,\frac{e^{-i\rho x}\, dx}{\sinh^{2n}{(x-i\epsilon)}}=\frac{(-1)^{n}}{(2n-1)!}\left(\frac{2\pi}{\rho}\right)\frac{1}{e^{\pi\rho}-1}\prod_{k=1}^{n}\left[\rho^{2}+4(n-k)^{2}\right]~.
\label{general_integral}
\end{eqnarray}
We use the above integral to obtain the integral of our concern, as in eq (\ref{fdt_de_1}). Therefore $I_1$ and $I_2$ turns out to be, 
\begin{align}
&I_1=\,-\f{1}{4!}\, \bigg(\f{2\pi}{\o'}\bigg)\, \f{\o'^2\,(\o'^2+4)\,(\o'^2+16)}{e^{\pi\,\o'}-1};~~~~~~~~~~I_2=\f{4}{3!}\, \bigg(\f{2\pi}{\o'}\bigg)\, \f{\o'^2\,(\o'^2+4)}{e^{\pi\,\o'}-1}~.
\label{I_desitter}
\end{align}
Here, $\o'=2\o\,\a_d$. Using the above expression for $I_1$ and $I_2$, we obtain $K(-\o)$ and $K(\o)$ as following,
\beq
K(-\o)=\,\f{\o^3\,(\a_{d}^{2}\o^2+1)}{12\a_{d}^{2}\pi\,\bigg(e^{2\pi\,\a_{d}\o}-1\bigg)};~~~~~~~~~K(\o)=\,\f{(-1)\o^3\,(\a_{d}^{2}\o^2+1)}{12\a_{d}^{2}\pi\,\bigg(e^{-2\pi\,\a_d\o}-1\bigg)}~.
\label{K_desitter_2}
\eeq
In our parametrisation the symmetric and anti-symmetric correlation functions can be written as, 
\begin{eqnarray}
K^{+}(\omega)= K(\omega)+K(-\omega)\:\:\:\:\:\:\text{and}\:\:\:\:\:\: K^{-}(\omega)=K(\omega)-K(-\omega)~.
\end{eqnarray}
Therefore, using eq. (\ref{K_desitter_2}), the ratio of the symmetric and the antisymmetric correlation functions turn out to be as following, 
\begin{eqnarray}
\frac{K^{+}(\omega)}{K^{-}(\omega)}=\coth\bigg({\frac{2\pi\a_d\omega}{2}}\bigg)~. 
\label{fdt_bh}
\end{eqnarray}

\subsubsection{$(3+1)$ dimensions}
We proceed similarly in case of the $(3+1)$-dimensional FLRW spacetime, and obtain $K(-\o)$ and $K(\o)$ as following,
\begin{align}
K(-\o)&=\f{1}{6\,\a_{d}^{6}\,\pi^3}\,\f{\o^3\,(\a_{d}^{2}\o^2+1)\,(\a_{d}^{2}\o^2+4)\,(\a_{d}^{2}\o^2+9)}{8!\,\bigg(e^{2\pi\,\a_{d}\o}-1\bigg)}~;
\nonumber\\
K(\o)&=\f{(-1)}{6\,\a_{d}^{6}\,\pi^3}\,\f{\o^3\,(\a_{d}^{2}\o^2+1)\,(\a_{d}^{2}\o^2+4)\,(\a_{d}^{2}\o^2+9)}{8!\,\bigg(e^{-2\pi\,\a_{d}\o}-1\bigg)}~.
\label{K_desitter_4}
\end{align}
Using the above set of equations one obtains the corresponding $K^{+}(\omega)$, $K^{-}(\omega)$ and their ratio, which in turn satisfies the relation as described in eq. (\ref{fdt_bh}). 

This implies that a comoving observer in (1+1) and as well as in (3+1) dimensional FLRW spacetime observes that the fluctuations of the random force, produced in the conformal vacuum of the massless scalar field, abide by the FDT. The temperature of the thermal bath of the produced particles, as detected by the comoving observer is correctly identified by comparing with eq. (\ref{fdt_1}) as $\f{1}{2\pi \a_{d}}$. We discuss the physical implications of these findings in the upcoming section.

\subsection{($1+1$) dimensional Schwarzschild BH}
In case of the BH spacetime, we proceed similarly with the Fourier transformation of the correlation function (i.e, eq. (\ref{correlation_sch_k1}) which is same for Kruskal and Unruh vacua), corresponding to the Schwarzschild BH spacetime, as follows, 
\begin{align}
K(-\omega)=&\underbrace{(-A) \int_{-\infty}^{\infty}d\tau\: e^{-i\omega\tau}\times\frac{5}{\sinh^{6}{\left(\frac{\tau}{8M\sqrt{1-\frac{2M}{R}}}\right)}}}_{\mathcal{I}_{1}} \nonumber\\
&\qquad + \underbrace{(-A) \int_{-\infty}^{\infty}d\tau\: e^{-i\omega\tau}\times\frac{4}{\sinh^{4}{\left(\frac{\tau}{8M\sqrt{1-\frac{2M}{R}}}\right)}}}_{\mathcal{I}_{2}}~,
\label{ft_bh_1}
\end{align}
where $A=\frac{\left(1-\frac{2M}{R}\right)^{-3}}{2\: \pi^{2}\: (8\: M)^{6}}$.
Using the general expression as in eq.(\ref{general_integral}), we obtain,
\begin{eqnarray}
&&\mathcal{I}_{1}=\frac{5\,q\,A}{5!}\:\frac{2\pi}{\omega '}\:\frac{\omega '^{2}}{e^{\pi\:\omega '}-1}(\omega '^{2}+4)(\omega '^{2}+16)~;
\label{ft_1}\\
\text{and}
\nonumber\\
&&\mathcal{I}_{2}=-\frac{4\,q\,A}{3!}\:\frac{2\pi}{\omega '}\:\frac{\omega '^{2}}{e^{\pi\:\omega '}-1}(\omega '^{2}+4)~,
\label{ft_2}
\end{eqnarray}
where we have defined, 
$\omega '\equiv q\omega$, and, $q\equiv8M\sqrt{1-\frac{2M}{R}}$.~
Using, $\mathcal{I}_1,\,\mathcal{I}_2$ from eqs. (\ref{ft_1}) and (\ref{ft_2}) in eq.(\ref{ft_bh_1}), we obtain,
\begin{eqnarray}
K(-\omega)= \frac{A\,q^4\,\pi}{12}\:\frac{\omega^{3}(q^2\omega^{2}+4)}{e^{q\pi\omega}-1};\qquad K(+\omega)=  -\,\frac{A\,q^4\,\pi}{12}\:\frac{\omega^{3}(q^2\omega^{2}+4)}{e^{-q\pi\omega}-1}~.
\end{eqnarray}
Therefore, following the same method as we have implemented in case of the FLRW spacetime, the ratio of the symmetric and the antisymmetric correlation functions turn out to satisfy the relation, as described in eq. (\ref{fdt_1}).  
This signifies that from the perspective of an observer in Schwarzschild spacetime the fluctuations of the random force, produced in the respective vacuum states such as Kruskal/Unruh vacuum, obeys FDT. 
From the above relation one can identify the temperature of the thermal bath of the produced particles from the corresponding vacuum states as, 
\begin{eqnarray}
T=\frac{1}{8\pi M\sqrt{1-\frac{2M}{R}}}~.
\end{eqnarray}
This exactly matches with the result obtained by the detector response method, as mentioned earlier.
\section{Finding the dissipative coefficient}\label{dissipation}
So far we consider the fluctuating part of the random force and after obtaining the correlation function of the fluctuations of the random force, one can in principle also obtain the dissipative coefficient of the corresponding dissipative force. This correlation between the fluctuating part and the dissipative component of the system in thermal equilibrium is another way of stating the FDT. It is evident that in general, the dissipative force will be retarded and therefore the general time-dependent dissipative coefficient $(\gamma(t))$ in the frequency domain can be written as \cite{kubo_book},
\be
{\rm Re}\{\gamma[\o]\}=\f{1}{2kT}\int_{-\infty}^{\infty}\langle R^{\s}(0)\,R^{\s}(\ta)\rangle \,e^{-i\o t} dt~.
\label{dissi_coeff}
\ee
 We use the derived expression for $K(-\o)$ for $(1+1)$ and $(3+1)$-dimensional de Sitter FLRW Universe,
 as these two integrals are similar in structure, and obtain the real dissipative coefficient in frequency domain respectively as,
\be
\gamma(\o)_{(2)}=\,\f{\o^{3}\,(\a_{d}^{2}\o^{2}+1)}{24\,kT\,\a^{2}_{d}\,\pi(e^{2\pi\a_d\o}-1)}~;
\label{dissi_coeff_2}
\ee
\be
\gamma(\o)_{(4)}=\f{1}{12\,kT}\,\f{\o^{3}\,(\a_{d}^{2}\o^{2}+1)(\a_{d}^{2}\o^{2}+4)(\a_{d}^{2}\o^{2}+9)}{8!\,\a^{6}_{d}\pi^{3}\,(e^{2\pi\a_d\o}-1)}~.
\label{dissi_coeff_4}
\ee
Here $\gamma_{(2)}$ and $\gamma_{(4)}$ symbolise the real dissipative coefficients in frequency space in $(1+1)$, $(3+1)$-dimensional de Sitter FLRW Universe respectively.
 
Similarly in the context of Schwarzschild spacetime the dissipative coefficient in frequency domain can be found as,
\be
\gamma(\o)_{{\rm(sc)}}=\,\f{A\,q^4\,\pi\,\o^{3}\,(q^2\o^{2}+1)}{24\, kT\,(e^{q\pi\o}-1)}\label{dissi_coeff_1}
\ee
where, $A,\,q,\, \o'$ symbolise the same as defined earlier in Section \ref{fd_property} and $\gamma(\o)_{{\rm(sc)}}$ stands for the real dissipative coefficient in the frequency domain, evaluated in the background of the Schwarzschild spacetime.  
 The parameter $T$ in all these cases symbolises the temperature of the corresponding thermal bath of the produced particles as perceived by the privileged class of observer. 
 
\section{Conclusion and Discussions}\label{conclusion}
The background of this present work relies on the fact that the observer/detector plays an important role when it comes to defining the particle content in a state, as well as the precise meaning of the vacuum state of a quantum field in curved spacetime. We consider the simplest case of all, i.e, a massless quantum scalar field in curved spacetime with conformal symmetry such as FLRW and $(1+1)$-dimensional Schwarzschild spacetime and discuss the phenomena of particle production from the suitable (though not unique) choice of vacuum states, with respect to some privileged class of observers. Due to the production of these particles, the same class of observers perceive the thermal spectrum from the corresponding vacuum states. In this context, we address a question from the perspective of these observers: what would be the impact of these produced particles on some test particles, which are released in the thermal bath of these produced quanta? Due to the involvement of a large number of particles, we adapt the techniques from statistical field theory, where we aim to obtain the correlation function of the fluctuations of the random force as applied by the produced quanta on the test particles, instead of calculating a large number of force equations for each produced particles. We formulate the random force, exerted by these produced particles and subsequently the fluctuations in the random force also have been calculated. 

We determine the correlation function of the fluctuating part of the random force, evaluated with respect to the corresponding vacuum states, from the perspective of a privileged class of observers. Our result depicts that to these observers, the behaviour of the entire system of the produced particles and as well as the test particles resembles with the systems in nonequilibrium statistics. The observer measures that the correlation function of the fluctuations of the random force is obeying the FDT. As a consequence, the observer will perceive that the test particles are executing Brownian-like motion in that thermal bath. We show that this feature is similar in all curved spacetimes of our concern in this present work irrespective of the dimensionality of the spacetime. In case of the $(3+1)$-dimensional FLRW spacetime, due to the isotropic and the homogeneous nature of the spacetime, the correlation function of the fluctuations of the random force will be identical in all directions. Also, the $(1+1)$-dimensional Schwarzschild BH case for a static observer in Kruskal and Unruh vacua has been investigated, which also led to a similar conclusion.  

We mention that in principle, one could make an attempt to write down the Langevin equation for these kinds of systems. However, the lack of information regarding the velocity profile of the test particles, due to the applied force of the produced particles prevents us from formulating the Langevin equation for such systems. The coefficient of the mean dissipative force (as obtained in section \ref{dissipation}) may help us to construct the dissipative part of the force equation and formulate the Langevin equation, but that needs deeper investigation. In order to fulfil that purpose, we need to incorporate all the forces arising in the system. In this context, some works can be mentioned here. In \cite{Unruh:1989} the authors have considered the test particle as an accelerating quantum harmonic oscillator which is interacting with the massless scalar field in the Minkowski spacetime. The form of the interaction term is chosen to be proportional to the velocity of the field. In this model, people first found the Langevin equation and then they studied the fluctuation-dissipation of the system (also see the subsequent papers \cite{Raine:1991,Hinterleitner:1993,Kim:1997,Kim:1998,Massar:1993,Massar:2006} for further progress).

We have mentioned earlier that obtaining the Langevin equation for our system by analysing the dissipative coefficient, would be a possible future work. In this context, another possible future direction would be to extend the present analysis for the radiation and matter-dominated era of the Universe, where the corresponding positive frequency Wightman functions lose their time translational invariance. In the present work, we confine ourselves to the production of the scalar quanta, whereas an important extension would be to analyse the production of other quantum fields such as fermions and gauge bosons and study their respective behaviour from the perspective of the privileged class of observers, in different curved spacetimes. Another prospective would be to examine the characteristic behaviour of produced quanta  in the context of the other possible $(1+1)$ dimensional BH solutions such as discussed in \cite{Lemos:1994fn, Ref2, Ref3}. BH solutions as in \cite{Lemos:1994fn, Ref2, Ref3} are described in $(1+1)$ dimensional spacetime and therefore can be expressed as the conformally flat spacetime. Subsequently one can follow the same procedure, as adapted in this present manuscript, and conclude accordingly.

So far we observed that a test particle in the thermal bath, seen from some particular frames, exhibit random motion which is Brownian in nature. This implies that each of the produced particles itself will also follow the same law due to the force exerted by the others in the thermal bath (as each particle in the bath can be regarded as a test particle). Consequently, every produced particle will also exhibit random motion consistent with FDT. Another point will be worth mentioning. It is well known that near horizon, the BH spacetimes are effectively $(1+1)$-dimensional \cite{Carlip:1998wz, Robinson:2005pd, Iso:2006wa, Majhi:2011yi}. Note that particle production phenomenon for a BH is very near horizon event and also we observed that for $(1+1)$-dimensional BH the static observer will see FDT in Kruskal and Unruh vacuums. Therefore, as far as the near horizon is concerned, the same will also happen for higher-dimensional black holes. This a suggestive statement, rather than a conclusive one. Of course, to reach a definite conclusion, a rigorous analysis has to be done by taking into account the other transverse dimensions.

\section*{Appendices}
\appendix
\section{Expectation value of the component $T_{t}\,^{t}$ for de Sitter Universe}\label{app1}
The present work is based on the phenomena of the particle production in the conformal vacuum of the quantum field, with respect to the particular class of observer in de Sitter FLRW and Schwarzschild BH spacetime. Therefore the expectation value of the $T_{tt}$ component of the stress-energy tensor for the corresponding field, evaluated with respect to the conformal vacuum should turn out to be nonzero.  In this section we aim to find the renormalised expectation value of the $T_{tt}$ component of the scalar field with respect to the conformal vacuum in $(1+1)$ and as well as $(3+1)$-dimensional FLRW spacetime (for massive scalar in FLRW, see \cite{Parker:1974}). Subsequently, we find the same in the background of BH spacetime in the Appendix \ref{app3}. 

\subsection{$(1+1)$-dimensional FLRW spacetime}
It is described earlier that the two-dimensional curved spacetime can be written as the conformally flat spacetime which is depicted in eq. (\ref{conformal_min}). It is well described in \cite{Birrell}, that due to this fact it is possible to write down the expectation value of the stress tensor in $(1+1)$-dimensional curved spacetime, in terms of the expectation value of the stress tensor in flat spacetime. 
To proceed further, following the discussion in \cite{Birrell}, we switch to the null coordinate system in order to represent the $(1+1)$-dimensional curved spacetime which is conformally flat to the Minkowski spacetime, as following, 
\beq
ds^{2}_{(N)}=C(u,v)\,du\,dv~.
\label{null_1}
\eeq 
Here, $(u,v)$ symbolise the corresponding null coordinates system. In the above metric background, the renormalised expectation value of the stress tensor components, can be written as (see Eq. (6.136) of \cite{Birrell}),
\beq
\braket{T_{a}\,^{b}[g_{c d}(x)]}_{\rm{ren}}=\frac{\sqrt{-\eta}}{\sqrt{-g}}\braket{T_{a}\,^{b}[\eta_{c d}(x)]}_{\rm{ren}}+\theta_{a}\,^{b}-\f{1}{48\pi}\,\mathcal{R}^{(2)}\,\delta_{a}\,^{b}~,
\label{ren_em_1}
\eeq
where, 
\beq
&\theta_{uu}\,=\,-\f{1}{12\pi}\,C^{1/2}\,\p_{u}^2\,\bigg(C^{-1/2}\bigg)~;\nonumber\\
&\theta_{vv}\,=\,-\f{1}{12\pi}\,C^{1/2}\,\p_{v}^2\,\bigg(C^{-1/2}\bigg)~;\nonumber\\
&\theta_{uv}\,=\,\theta_{vu}\,=\,0~.
\label{thetas_1}
\eeq
In the above, $(a,b)$ (i.e, the alphabets in lower case), stand for all the spacetime coordinates $(t,x)$. In this present context $(a,b)$ symbolise the null coordinates $(u,v)$. In our case $\sqrt{-\eta}$, corresponds to the determinant of the flat spacetime in the null coordinate $(u,v)$. Therefore one finds $\eta_{uv}=\eta_{vu}=\f{1}{2}$. $\mathcal{R}^{(2)}$ is the $(1+1)$-dimensional Ricci scalar corresponding to the curved spacetime. In the context of $(1+1)$-dimensional FLRW spacetime, the quantity $\braket{T_{a}\,^{b}[g_{cd}(x)]}_{\rm{ren}}$ symbolises the renormalised expectation value of the stress energy tensor for the conformally flat metric, evaluated with respect to the conformal vacuum. We mention that this quantity is indeed measured by the comoving observer. Similarly, $\braket{T_{a}\,^{b}[\eta_{cd}(x)]}_{\rm{ren}}$ signifies the renormalised expectation value of the stress energy tensor, corresponding to the flat spacetime part of the full metric (\ref{null_1}) as measured by the comoving observer. Regarding the specification of the vacuum state of the later, it is well described in \cite{Birrell}, that if the conformal spacetime is conformal to the whole Minkowski spacetime (i.e, not just only a part of the Minkowski spacetime), the usual Minkowski vacuum state is employed to calculate the expectation value of the stress tensor. Under this circumstances the quantity, $\braket{T_{a}\,^{b}[\eta_{cd}(x)]}_{\rm{ren}}$ will turn out to be zero. However if the vacuum state is not the Minkowski vacuum state, the first term on the right hand side of eq. (\ref{ren_em_1}), will produce a nonzero contribution. 
In the above equations, $C$ is related with the conformal factor $\O^2(x)$, which appears in eq. (\ref{conformal_min}) in Section \ref{description_frw}, as, $\O^{2}(x)=C$. For $(1+1)$-dimensional FLRW spacetime, $(u,v)$ can be depicted as, $u=(\eta-x)$, $v=(\eta+x)$, by following the metric as in eq. (\ref{frw_2d}). In this case, the FLRW spacetime, in terms of the null coordinates, can be written as in eq. (\ref{null_1}), where $C(u,v)=\f{2\alpha_{d}^{2}}{(u+v)^2}\,du\,dv$. 

Our quantity of interest is the renormalised expectation value of the components of the stress energy tensors, evaluated with respect to the conformal vacuum and measured by a comoving observer i.e, $\braket{T_{t}\,^{t}}_{\rm{ren}}$. Therefore at first to implement eq. (\ref{ren_em_1}), we transform tensor component $T_{t}\,^{t}$ in terms of the null coordinates $(u,v)$, by following the step as described in Section \ref{mathematical_analysis}. Hence we obtain,
\beq
\bra{0}T_{t}\,^{t}\ket{0}_{\rm{ren}}=\f{1}{2}\,\bra{0}T_{u}\,^{v}+T_{v}\,^{u}+2T_{u}\,^{u}\ket{0}_{\rm{ren}}~.
\label{ren_frw_tt_1}
\eeq
Now we write $T_{u}\,^{v},\,T_{v}\,^{u},\,T_{u}\,^{u}$, which are defined in curved spacetime, by following eq. (\ref{ren_em_1}), where we take $(a,b)=(u,v),\,(v,u),\,(u,u)$ respectively. For example,
\beq
\f{1}{2}\braket{T_{u}\,^{v}[g_{cd}(x)]}=\f{\sqrt{-\eta}}{2\sqrt{-g}}\braket{T_{u}\,^{v}[\eta_{c d}(x)]}_{\rm{ren}}+\f{1}{2}\theta_{u}\,^{v}~,
\label{ren_frw_tt_2}
\eeq
Here one finds $\sqrt{-\eta}=\f{1}{2}$ and for the present case, $(c,d)$ correspond to the null coordinates. In case of $(1+1)$-dimensional FLRW, the conformal vacuum is same as the usual Minkowski vacuum and therefore the first term on the right hand side of eq. (\ref{ren_frw_tt_2}) vanishes. In order to show this explicitly, one can indeed start the analysis by writing the general form of the stress tensor $T_{u}\,^{v}[\eta_{c d}(x)]$, in terms of the field $\phi(x)$ as following,  
\begin{align}
T_{ab}^{(\phi)}(x)&=(1-2\xi)\nabla_{a}\phi\,\nabla_{b}\phi+\bigg(2\xi-\f{1}{2}\bigg)\,g_{ab}\,g^{cd}\,\nabla_{c}\phi\,\nabla_{d}\phi-2\xi\,[\nabla_{a}\nabla_{b}\,\phi]\phi \nonumber \\
&\quad + \f{2}{D}\xi\,g_{ab}(\phi\,\Box{\phi})-\xi\bigg[G_{ab}+\f{2(D-1)}{D}\,\xi\,Rg_{ab}\bigg]\,\phi^{2}+2\bigg[\f{1}{4}-\left(1-\f{1}{D}\right)
\xi\bigg]\,m^2\,g_{ab}\phi^{2}~,
\label{gen_em_1}
\end{align}
where $D$ symbolises the dimensionality of spacetime and $\xi$ is described in Section \ref{description_frw}.
It is mentioned earlier that in this work we consider the scalar field to be massless and therefore the last term in eq. (\ref{gen_em_1}) will not contribute further. 
In two-dimensional spacetime $\xi$ turned out to be zero and produces, $T_{ab}(x) \sim \partial_{a}\phi\,\partial_{b}\phi$. Upon using the point splitting technique, one can reduce the expectation value of the stress tensor in terms of the derivatives of the positive frequency Wightman function. As both the conformal and the Minkowski vacuum states signify the same vacuum state, therefore the expectation value i.e, $\braket{T_{u}\,^{v}[\eta_{c d}(x)]}_{\rm{ren}}$ will turn out to be zero in this case (the point splitting technique will be described elaborately in case of the $(1+1)$-dimensional BH spacetime, in Section \ref{app3}, as it produces a non zero result for this quantity).

For $(1+1)$-dimensional de Sitter FLRW we also have $\theta_{u}\,^{v}=g^{uv}\,\theta_{uu}$. It can be noted that for $C^{-1/2}=\f{(u+v)}{\sqrt{2}\alpha_d}$, $\theta_{uu}=\theta_{vv}=0$. Therefore $\braket{T_{u}\,^{v}[g_{cd}(x)]}=0$. Proceeding similarly the second term of eq. (\ref{ren_frw_tt_1}), also becomes zero. The only contribution comes from the third term of eq. (\ref{ren_frw_tt_1}), which is 
\beq
\braket{T_{u}\,^{u}[g_{cd}(x)]}_{\rm{ren}}=\f{\sqrt{-\eta}}{\sqrt{-g}}\braket{T_{u}\,^{u}[\eta_{c d}(x)]}_{\rm{ren}}+\theta_{u}\,^{u}-\f{1}{48\pi}\,\mathcal{R}^{(2)}~.
\label{ren_em_2}
\eeq
By the same reason as explained in case of eq.(\ref{ren_frw_tt_2}), the first term on the right hand side of the above equation turns out to be zero. The second term of the above equation, $\theta_{u}\,^{u}=g^{uv}\,\theta_{uv}=0$. 
Therefore using the Ricci scalar $\mathcal{R}^{(2)}=-\f{2}{\alpha_{d}^{2}}$ and $C(u,v)$ corresponding to the $(1+1)$-dimensional de Sitter FLRW spacetime, in the above equation, we finally obtain,
\beq
\braket{T_{t}\,^{t}[g_{cd}(x)]}_{\rm{ren}}=\f{1}{24\pi\alpha_{d}^{2}}~.
\label{ren_em_3}
\eeq
This is the correct expression for energy flux which is a non-zero constant value.

\subsection{(3+1) Dimensions}\label{renormalisation_3_frw}
In the $(3+1)$-dimensional FLRW spacetime, the renormalised expectation value of the $T_{tt}$ component of the stress tensor, as evaluated by a comoving observer with respect to the conformal vacuum of the massless scalar field, can be written in terms of the curvature of the corresponding curved spacetime as follows (see Eq. (7.44) of \cite{Birrell}), 
\beq
\bra{0}T_{ab}\ket{0}_{\rm{ren}}=&\f{1}{2880\,\pi^2}\,\bigg[\bigg(-\f{1}{3}\,\nabla_{a}\nabla_{b}\mathcal{R}^{(4)}+\mathcal{R}_{a}\,^{c(4)}\,\mathcal{R}_{cb}^{(4)}-\mathcal{R}^{(4)}\,\mathcal{R}_{ab}^{(4)}\bigg)\nonumber\\
&+g_{ab}\,\bigg(\f{1}{3}\,\Box \mathcal{R}^{(4)}-\f{1}{2}\mathcal{R}^{cd(4)}\,\mathcal{R}_{cd}^{(4)}+\f{1}{3}\mathcal{R}^{(4)2}\bigg)\bigg]~.
\label{ren_em_5}
\eeq
In the above equation $g_{ab}$ symbolises the metric tensor corresponding to the $(3+1)$-dimensional de Sitter FLRW spacetime and $\mathcal{R}^{(4)},\,\mathcal{R}_{ab}^{(4)}$ is the Ricci scalar, Ricci tensor, defined with respect to $g_{ab}$ in $(3+1)$ dimensions. In this spacetime the Ricci scalar turns out to be: $\mathcal{R}^{(4)}=-\f{12}{\alpha_{d}^{2}}$ and for the Ricci tensor, only the diagonal components survive. Surviving components of Ricci tensor are as follows, $\mathcal{R}_{xx}^{(4)}=\mathcal{R}_{yy}^{(4)}=\mathcal{R}_{zz}^{(4)}=\f{3}{\alpha_{d}^{2}}e^{\f{2t}{\alpha_d}}$, and $\mathcal{R}_{tt}^{(4)}=-\f{3}{\alpha_{d}^{2}}$. Upon using these expressions, we obtain the renormalised expectation value of the $T_{tt}$ component as, 
\beq
\bra{0}T_{t}\,^{t}\ket{0}_{\rm{ren}}=\,\f{1}{960\pi^2\alpha_{d}^{4}}~,
\eeq
which is again a constant.
This result depicts that the comoving observer in $(3+1)$-dimensional de Sitter FLRW spacetime, will effectively perceive a finite expectation value for the $(t,t)$ component of the stress-energy tensor, which signifies the presence of a finite energy density of the produced scalar field quanta in the conformal vacuum of the scalar field. 

\section{Expectation value of the component $T_{t}\,^{t}(t,r)$ for the BH spacetime}\label{app3}
In this section we calculate the renormalised expectation value of the $T_{t}\,^{t}(t,r)$ component of the stress energy tensor in the background of the Schwarzschild spacetime. We follow the same procedure as adapted for the $(1+1)$-dimensional FLRW spacetime, described in the previous section.
We consider the case where the Schwarzschild observer detects the particle production in the Unruh vacuum. Like earlier cases of FLRW spacetime, here also we aim to calculate the renormalised expectation value of the $(t,t)$ component of the stress-energy tensor as evaluated by the Schwarzschild observer with respect to the Unruh vacuum/\,Kruskal vacuum. In this case, the Unruh/\,Kruskal vacuum state does not coincide with the Minkowski vacuum, which can be portrayed as the Boulware vacuum. Hence the quantity $\braket{T_{a}\,^{b}[\eta_{cd}(x)]}_{\rm{ren}}$ in eq. (\ref{ren_em_1}), produces a nonzero contribution and has to be evaluated separately. 

This quantity is nothing but the difference between the expectation value of the stress tensor components evaluated with respect to the Unruh/\,Kruskal vacuum and Minkowski vacuum, measured by the Schwarzschild observer. 
Proceeding similarly as $(1+1)$-dimensional FLRW spacetime, we obtain,
\beq
\braket{T_{t}\,^{t}}_{\rm{ren}}=\frac{1}{2}\bra{0}T_{u}\,^{v}+T_{v}\,^{u}+2T_{u}\,^{u}\ket{0}_{\rm{ren}}~.
\label{ren_sc_tt_1}
\eeq
In case of the BH spacetime the null coordinates $(u,v)$ are defined in the Section \ref{examples}. We start our analysis by considering the first term on the right hand side of eq. (\ref{ren_sc_tt_1}), evaluated with respect to the Unruh vacuum. Later, we generalise this same procedure for the Kruskal vacuum case. Upon using eq. (\ref{ren_em_1}), we obtain,
\beq
\f{1}{2}\braket{T_{u}\,^{v}[g_{cd}(x)]}_{\rm{ren(U)}}&=&\f{\sqrt{-\eta}}{2\,\sqrt{-g}}\bigg[\,_{U}\langle 0|T_{u}\,^{v}[\eta_{cd}(x)]|0\rangle_{U}\,-\,_{M}\langle 0|T_{u}\,^{v}[\eta_{cd}(x)]|0\rangle_{M}\bigg]+\f{1}{2}\theta_{u}\,^{v}\nonumber\\
&=&\f{\eta^{uv}\sqrt{-\eta}}{2\,\sqrt{-g}}\bigg[\,_{U}\bra{0}T_{uu}[\eta_{cd}(x)]\ket{0}_U-_{M}\bra{0}T_{uu}[\eta_{cd}(x)]\ket{0}_M \bigg]+\f{1}{2}\theta_{u}\,^{v}~.\nonumber\\
&=&\bigg(1-\f{r_s}{r}\bigg)^{-1}\bigg[\,_{U}\bra{0}T_{uu}[\eta_{cd}(x)]\ket{0}_U-_{M}\bra{0}T_{uu}[\eta_{cd}(x)]\ket{0}_M \bigg]+\f{1}{2}\theta_{u}\,^{v}~.
\nonumber\\
\label{ren_sc_tt_3}
\eeq
Here we write, $\braket{T_{u}\,^{v}[\eta_{cd}(x)]}_{\rm{ren(U)}}=\bigg[\,_{U}\langle 0|T_{u}\,^{v}[\eta_{cd}(x)]|0\rangle_{U}\,-\,_{M}\langle 0|T_{u}\,^{v}[\eta_{cd}(x)]|0\rangle_{M}\bigg]$. In this context we would like to mention that in \cite{Candelas:1976,Candelas:1977,Dowker:1978}, a similar term was calculated for an accelerating frame in four-dimensional flat spacetime. There the same has been interpreted as the components of energy-momentum tensor for an accelerating plane conductor. This can be also called as {\it vacuum stress} \cite{Takagi:1986kn}. In our present case, we call this as the stress of the Unruh vacuum. We use the explicit form of the $T_{uu}$ in terms of the scalar field in $(1+1)$-dimensional spacetime by following the eq.(\ref{gen_em_1}) and implement the point splitting technique. Therefore the first term on the right hand side of the above equation reduces to,
\beq
\bigg(1-\f{r_s}{r}\bigg)^{-1}\bigg[\lim_{x\to x'}\p_{u}\partial_{u'}\,_{U}\bra{0}\phi(u,v)\phi(u',v')\ket{0}_U
-\lim_{x\to x'}\p_{u}\partial_{u'}\,_{M}\bra{0}\phi(u,v)\phi(u',v')\ket{0}_M\bigg]~.\nonumber\\
\label{ren_sc_tt_2}
\eeq
Here we denoted $(x;x')=(u,v;\,u',v')$. The positive frequency Wightman function of a scalar field, evaluated with respect to the Unruh vacuum and as measured by the Unruh observer,  is well known in literature (depicted in eq.(\ref{green_unruh})). Hence to evaluate the term within the square bracket, in the above expression, at first we use eq. (\ref{green_unruh}) and then transform the coordinates $(\bar{u},v)\to \,(u,v)$. Performing the derivatives, we set the trajectory of the Schwarzschild observer to be $r=r'=\,R$ (i.e, constant curvature). Therefore, the above term reduces to,
\beq
\bigg(1-\f{r_s}{r}\bigg)^{-1}\bigg(-\f{1}{4\pi}\bigg)\bigg[\lim_{x\to x'}\bigg(\f{1}{64\,M^2}\bigg)\f{1}{\rm{sinh}^2(\f{\Delta u}{8M})}-\lim_{x\to x'}\f{1}{(\Delta u)^2}\bigg]
\eeq
where $\Delta u= (u-u')$. The second term within the square braket of the above equation, corresponds to the expectation value, evaluated with respect to the Minkowski vacuum and hence, by following the same point splitting method, it reduces to,  
\beq
\lim_{x\to x'}\p_{u}\partial_{u'}\,\bigg[\,_{M}\bra{0}\phi(u,v)\phi(u',v')\ket{0}_M\bigg]=\bigg(-\f{1}{4\pi}\bigg)\,\lim_{\Delta u\to \,0}\bigg[\f{1}{(\Delta u)^2}\bigg]~.
\eeq
Subsequently, using the series expansion of ${\rm sinh}(\f{\Delta u}{2\a_{d}})$, eq. (\ref{ren_sc_tt_2}) reduces to,
\beq
\bigg(1-\f{r_s}{r}\bigg)^{-1}\bigg(-\f{1}{4\pi}\bigg)&&\lim_{\Delta u\to 0}\bigg[\f{1}{64M^2}\bigg\{\bigg(\f{8M}{\Delta u}\bigg)^2\bigg(1+\f{1}{3!}\bigg(\f{\Delta u}{8M}\bigg)^2+\f{1}{5!}\bigg(\f{\Delta u}{8M}\bigg)^4+...\bigg)^{-2}\bigg\}
\nonumber
\\
&&-\f{1}{(\Delta u)^2}\bigg]~.
\label{ren_sch_15}
\eeq
Upon further simplification the above expression becomes,
\beq
\bigg(1-\f{r_s}{r}\bigg)^{-1}\bigg(-\f{1}{4\pi}\bigg)\lim_{\Delta u \to 0}\bigg[\f{1}{(\Delta u)^2}-\underbrace{\f{1}{192 M^2}}_{\rm{finite\, term}}-\f{1}{(8M)^4}\f{(\Delta u)^2}{60}-.....-\f{1}{(\Delta u)^2}\bigg]~.
\label{ren_sch_4}
\eeq
Implementing $\Delta u \to 0$, we are left with only the finite term as following,
\beq
\bigg(1-\f{r_s}{r}\bigg)^{-1}\f{1}{4\pi}\,\f{1}{192M^2}~.
\label{ren_em_6}
\eeq
Therefore we obtain eq. (\ref{ren_sc_tt_3}) as,
\beq
\f{1}{2}\braket{T_{u}\,^{v}[g_{cd}(x)]}_{\rm{ren(U)}}=\bigg(1-\f{r_s}{R}\bigg)^{-1}\f{1}{768\pi M^2}+\f{1}{2}\theta_{u}\,^{v}~.
\label{ren_em_9}
\eeq
Now proceeding similarly the second term on the right side of the eq. (\ref{ren_sc_tt_1}) becomes,
\beq
\f{1}{2}\braket{T_{v}\,^{u}[g_{cd}(x)]}_{\rm{ren(U)}}=\f{1}{2}\theta_{v}\,^{u}~.
\label{ren_em_10}
\eeq
Here, the term $\braket{T_{v}\,^{u}[\eta_{cd}(x)]}_{\rm{ren(U)}}$ reduces to zero. The third term of eq. (\ref{ren_sc_tt_1}) becomes,
\beq
\braket{T_{u}\,^{u}[g_{cd}(x)]}_{\rm{ren(U)}}=-\f{1}{48\pi}\mathcal{R}^{(2)}|_{r=R}=\f{r_s}{24\pi R^3}~.
\label{ren_em_11}
\eeq
Here also one obtains, $\braket{T_{u}\,^{u}[\eta_{cd}(x)]}_{\rm{ren(U)}}=0$ and $\theta_{u}\,^{u}=0$. We also use 
the two-dimensional Ricci scalar corresponding to the Schwarzschild spacetime as, $\mathcal{R}^{(2)}=-\f{2r_s}{r^{3}}$, where $r_s$ is specified in the earlier section.
Now we combine the terms $\theta_{u}\,^{v}$ and $\theta_{v}\,^{u}$ as appear in eqs. (\ref{ren_em_9}) and (\ref{ren_em_10}). Proceeding similarly as in the $(1+1)$-dimensional FLRW case,  we obtain 
\beq
\frac{1}{2}(\theta_{u}\,^{v}+\theta_{v}\,^{u})|_{r=R}=\f{1}{48\pi}\bigg[-\f{2r_s}{R^3}-\f{r_{s}^{2}}{2R^4}\bigg(1-\f{r_s}{R}\bigg)^{-1}\bigg]~.
\label{ren_em_7}
\eeq
In order to obtain the above equation we used, $C(r)=\f{1}{2}\bigg(1-\f{r_s}{r}\bigg)$. Now 
using the equations (\ref{ren_em_9})-(\ref{ren_em_7}) in eq. (\ref{ren_sc_tt_1}), we obtain the renormalised expectation value of the stress tensor as evaluated by the Schwarzschild observer with respect to the Unruh vacuum as,
\beq
\braket{T_{t}\,^{t}[g_{cd}(x)]}_{\rm{ren(U)}}=\bigg(1-\f{r_s}{R}\bigg)^{-1}\f{1}{768 \pi M^2}-\f{r_{s}^{2}}{96\pi\,R^4}\bigg(1-\f{r_s}{R}\bigg)^{-1}
\label{ren_em_8}
\eeq
Following the identical procedure, one can indeed obtain the renormalised expectation value of the stress tensor as evaluated by the Schwarzschild observer with respect to the Kruskal vacuum as,
\beq
\braket{T_{t}\,^{t}[g_{cd}(x)]}_{\rm{ren(K)}}=\bigg(1-\f{r_s}{R}\bigg)^{-1}\f{1}{384\pi M^2}-\f{r_{s}^{2}}{96\pi\,R^4}\bigg(1-\f{r_s}{R}\bigg)^{-1}~.
\label{ren_em_kruskal}
\eeq

Note that if the observer is at infinity ; i.e. $R\rightarrow\infty$, then (\ref{ren_em_8}) reduces to the well known Hawking expression $(\kappa^2/48\pi)$, with $\kappa = 1/4M$. Whereas the other one (\ref{ren_em_kruskal}) reduces to $(\kappa^2/24\pi)$ which, as expected, is two times of the Hawking expression. Also it can be checked that for Unruh vacuum, $T_{uu}$ vanishes which one expects. The explicit expression for this component is
\begin{equation}
\braket{T_{uu}[g_{cd}(x)]}_{\rm{ren(U)}} = \frac{\kappa^2}{48\pi} + \frac{1}{96\pi}\Big[-\frac{2r_s}{R^3}(1-\frac{r_s}{r})-\frac{r_s^2}{2R^4}\Big]~.
\label{Bibhas1}
\end{equation} 
Notice that the first term is constant, independent of $R$. If one solves the (trace) anomaly equation $T_a^a=\frac{\mathcal{R}^{(2)}}{24\pi}$ along with covariant conservation equation $\nabla_aT^{ab}=0$, the solution will be exactly identical to (\ref{Bibhas1}), where the first term comes as a integration constant (for instance, see Eq. (20) of \cite{Banerjee:2008sn}). This constant is fixed by a relevant boundary condition. In Unruh vacuum, the value of the constant is exactly the  same as appears in the above.
\section{Expectation value of the component $T_{t}\,^{\a}$ for de Sitter and Schwarzschild spacetime}\label{app2}
We compute the renormalised expectation value of the off diagonal component of the stress tensor, in the similar way as discussed in the earlier sections.

\subsection{de Sitter Universe}
{\it{$(1+1)$ dimensions}}: For the two-dimensional de Sitter FLRW spacetime, we use eq. (\ref{ren_em_1}), in order to evaluate the renormalised expectation value of the off diagonal component of the stress tensor i.e, $T_{t}\,^{x}$. By the tensor transformation the off diagonal component of the stress tensor $T_{t}\,^{x}$ can be written in terms of the null coordinates as, 
\beq
\braket{T_{t}\,^{x}[g_{cd}]}_{\rm{ren}}=\f{\eta}{2\,\alpha_d}\bigg[\braket{T_{v}\,^{u}[g_{cd}]}_{\rm{ren}}-\braket{T_{u}\,^{v}[g_{cd}]}_{\rm{ren}}\bigg]~.
\eeq
Here also, since the vacuum is conformal vacuum, we have $\braket{T_{v}\,^{u}[\eta_{cd}]}_{\rm{ren}}=\braket{T_{u}\,^{v}[\eta_{cd}]}_{\rm{ren}}=0$. Moreover we have explicitly shown earlier that in case of the two-dimensionalFLRW metric, $\theta_{uu}=\theta_{vv}=0$. In this case, the term associated with the Ricci scalar will also reduces to zero because of the presence of the Kronecker delta. Finally, the renormalised expectation value becomes, $
\braket{T_{t}\,^{x}[g_{cd}]}_{\rm{ren}}=0$.

{\it $(3+1)$ dimensions}: 
In $(3+1)$-dimensional FLRW spacetime we follow the procedure in the subsection \ref{renormalisation_3_frw}, and use eq. (\ref{ren_em_5}) in order to compute the renormalised expectation value of the off-diagonal components of the stress tensor as evaluated by the comoving observer with respect to the conformal vacuum. 
For the ease of the computation, we examine only the component $T_{t}\,^{x}(t,x)$. The outcome of the analysis for $T_{t}\,^{x}$ would be the same for other components (e.g. $T^y_t, T^z_t$) due to the homogeneity and isotopy of the FLRW spacetime. In this case as, $g_{tx}=0$, the second term on the right-hand side of the eq. (\ref{ren_em_5}) does not contribute to the expectation value. Moreover, it is mentioned earlier that in the background of the $(3+1)$-dimensional de Sitter FLRW spacetime, there exist only the diagonal components of the Ricci tensor. Hence terms like $\mathcal{R}_{t}\,^{x(4)}\mathcal{R}_{xx}^{(4)}$, $\mathcal{R}^{(4)}\mathcal{R}_{tx}^{(4)}$ become zero individually. Subsequently, one can clearly perceive that as the Ricci scalar corresponding to this spacetime, depends only on the expansion parameter $\alpha_d$, their derivatives are also going to be zero. Overall this analysis is implying that the renormalised expectation value of the off-diagonal components of the stress tensor as evaluated by the comoving observer with respect to the conformal vacuum in the $(3+1)$-dimensional de sitter FLRW spacetime, is turn out to be zero.

\subsection{two-dimensional BH}
In case of the BH spacetime, the quantity $\braket{T_{t}\,^{r}[g_{cd}(x)]}_{\rm{ren(U)}}$ is evaluated by following the discussion in the subsection (\ref{app3}). By the tensor transformation, the quantity $\braket{T_{t}\,^{r}[g_{cd}(x)]}_{\rm{ren(U)}}$ becomes,
\beq
\braket{T_{t}\,^{r}[g_{cd}]}_{\rm{ren(U)}}=\f{1}{2}\bigg(1-\f{r_s}{r}\bigg)\bigg[\braket{T_{u}\,^{v}[g_{cd}]}_{\rm{ren}}-\braket{T_{v}\,^{u}[g_{cd}]}_{\rm{ren}}\bigg]~.
\label{sch_offdiag_1}
\eeq
Following the same mathematical procedures one obtains,
\beq
\braket{T_{t}\,^{r}[g_{cd}]}_{\rm{ren(U)}}=\,\f{1}{768\pi M^2}~.
\label{sch_offdiag_2}
\eeq
Similarly, one can generalise the above procedures to the analysis of the expectation value of the off diagonal components of the stress tensor, evaluated with respect to the Kruskal vacuum, as measured by the Schwarzschild observer. This analysis produces,
\beq
\braket{T_{t}\,^{r}[g_{cd}]}_{\rm{ren(K)}}\,=\,0~.
\label{sch_offdiag_3}
\eeq
\section{Explicit derivation of the correlation function of the fluctuations of the random force}\label{explicit derivation_em}
The general form of the stress-energy tensor corresponding to the scalar field $\phi(x)$, is depicted in eq. (\ref{gen_em_1}). Following the method as described in subsection \ref{mathematical_analysis}, the term $\langle{0}|T^{tx}(t,\bar{x})T^{tx}(t',\bar{x'})|0\rangle$ in $(1+1)$ dimensions, i.e, eq. (\ref{correlation_flw_1}), can be written as,
\beq
\langle{0}|T^{tx}(t,\bar{x})T^{tx}(t',\bar{x'})|0\rangle=\f{1}{a^{3}(\eta)a^3(\eta')}\langle{0}|T_{\eta x}(\eta,\bar{x})T_{\eta x}(\eta',\bar{x'})|0\rangle~.
\label{random_app_1}
\eeq
In two-dimensionalspacetime $\xi$ turned out to be zero and we use $(a,b)=(\eta,x)$ in order to obtain the  $T_{\eta x}$ component of the stress energy tensor. This produces, $T_{\eta x}=\partial_{\eta}\phi\,\partial_{x}\phi$. Hence the stress-stress correlation function becomes, 
\beq
\langle{0}|T_{\eta x}(\eta,\bar{x})T_{\eta x}(\eta',\bar{x'})|0\rangle=\langle{0}|\partial_{\eta}\phi\,\partial_{x}\phi\,\partial_{\eta'}\phi\,\partial_{x'}\phi|0\rangle~.
\label{explicit_force_1}
\eeq
At this stage we use the Wick contraction and obtain, 
\begin{align}
\langle{0}|\partial_{\eta}\phi\,\partial_{x}\phi\,\partial_{\eta'}\phi\,\partial_{x'}\phi|0\rangle
&=\langle{0}|\partial_{\eta}\phi\,\partial_{\eta'}\phi|0\rangle\,\langle{0}|\partial_{x}\phi\,\partial_{x'}\phi|0\rangle+
\langle{0}|\partial_{\eta}\phi\,\partial_{x'}\phi|0\rangle\,\langle{0}|\partial_{x}\phi\,\partial_{\eta'}\phi|0\rangle\nonumber\\
&=\p_{\eta}\p_{\eta'}\bra{0}\phi(\eta,x)\phi(\eta',x')\ket{0}\,\p_{x}\p_{x'}\bra{0}\phi(\eta,x)\phi(\eta',x')\ket{0}\nonumber\\
&\qquad + \p_{\eta}\p_{x'}\bra{0}\phi(\eta,x)\phi(\eta',x')\ket{0}\,\p_{x}\p_{\eta'}\bra{0}\phi(\eta,x)\phi(\eta',x')\ket{0}\nonumber\\
&=\p_{\eta}\p_{\eta'}\bigg[G^{+}_{(2)}(\eta,x;\eta',x')\bigg]\,\p_{x}\p_{x'}\bigg[G^{+}_{(2)}(\eta,x;\eta',x')\bigg]\nonumber\\
&\qquad + \p_{\eta}\p_{x'}\bigg[G^{+}_{(2)}(\eta,x;\eta',x')\bigg]\,\p_{x}\p_{\eta'}\bigg[G^{+}_{(2)}(\eta,x;\eta',x')\bigg]~.
\label{explicit_force_2}
\end{align}
In the last step we use the positive frequency Wightman function $G^{+}_{(2)}(\eta,x;\eta',x')=\bra{0}\phi(\eta,x)\phi(\eta',x')\ket{0}$. After performing the derivatives which are appearing in the last step of the above equation, we need to put the trajectory of the comoving observer, which implies $\Delta x=0$ (proper frame condition). Therefore the second term of the last equation, reduces to zero and the final form of the stress stress correlation in $(1+1)$-dimensional spacetime becomes,
\beq
\langle{0}|T_{\eta x}(\eta,\bar{x})T_{\eta x}(\eta',\bar{x'})|0\rangle=\p_{\eta}\p_{\eta'}\bigg[G^{+}(\eta,x;\eta',x')\bigg]\,\p_{x}\p_{x'}\bigg[G^{+}(\eta,x;\eta',x')\bigg]~.
\label{explicit_force_3}
\eeq
This is nothing but the expression as used in eq. (\ref{correlation_flw_2}).

Proceeding similarly like the $(1+1)$-dimensional FLRW spacetime, the stress tensor component $T_{\eta x}$ in $(3+1)$ dimensions becomes,
\beq
T_{\eta x}=\f{2}{3}\p_{\eta}\phi\,\p_{x}\phi-\f{1}{3}\phi\,\p_{\eta}\p_x \phi+\f{a'(\eta)}{3a(\eta)}\phi\,\p_x \phi~.
\label{scalar_gen_4d}
\eeq
Implementing the Wick's contraction, the only survival terms we obtain for the stress-stress correlation function are,
\begin{align}
&\langle 0|T_{\eta x}(\eta, x) T_{\eta x}(\eta',x')|0\rangle \\
&=\f{4}{9}\,[\{\p_{\eta}\,\p_{\eta'}G^{+}(x,x')\}\,\{\p_{x}\p_{x'}G^{+}(x,x')\}]
\nonumber
\\
&\quad -\f{2}{9}\{\p_{\eta}G^{+}(x,x')\}\,\{\p_{x}\,\p_{\eta'}\,\p_{x'}G^{+}(x,x')\}\nonumber\\
&\qquad -\f{2}{9\eta'}\{\p_{\eta}G^{+}(x,x')\}\{\p_{x}\,\p_{x'}G^{+}(x,x')\}\}-\f{2}{9}\{\p_{\eta'}\{G^{+}(x,x')\}\,\{\p_{\eta}\p_{x}\p_{x'}G^{+}(x,x')\}\nonumber\\
&\qquad \quad+\f{1}{9}\{G^{+}(x,x')\}\{\p_{\eta}\p_{x}\p_{\eta'}\p_{x'}G^{+}(x,x')\}+\f{1}{9\eta'}\{G^{+}(x,x')\}\{\p_{\eta}\p_{x}\p_{x'}G^{+}(x,x')\}\nonumber\\
&\qquad \qquad -\f{2}{9\eta}\{\p_{\eta'}G^{+}(x,x')\}\{\p_{x}\p_{x'}G^{+}(x,x')\}+\f{1}{9\eta}\{G^{+}(x,x')\}\{\p_{x}\p_{\eta'}\p_{x'}G^{+}(x,x')\}\nonumber\\
&\qquad \qquad \quad +\f{1}{9\eta\eta'}\{G^{+}(x,x')\}\{\p_{x}\p_{x'}G^{+}(x,x')\}~,
\end{align}
where we use the positive frequency Wightman function in $(3+1)$-dimensional FLRW spacetime $G^{+}(x,x')=\bra{0}\phi(\eta,x)\phi(\eta',x')\ket{0}$, given by (\ref{mod_green_ds}). Now the expressions of the survival terms after performing the respective partial differentiations and implementing the proper frame condition ($\Delta x =0$) we obtain,
\begin{center}
\scalebox{1.3}{
\begin{tabular}{ ||c|c|| } 
 \hline
 $\f{4}{9}\,\{\p_{\eta}\,\p_{\eta'}G^{+}(x,x')\}\,\{\p_{x}\p_{x'}G^{+}(x,x')\}$ & $\f{\eta\eta'(\eta^{2}+4\eta\eta'+\eta'^{2})}{18\,\alpha_{d}^{4}\,\pi^{4}(\eta-\eta')^{8}}$\\ 
 \hline
 $-\f{2}{9}\{\p_{\eta}G^{+}(x,x')\}\,\{\p_{x}\,\p_{\eta'}\,\p_{x'}G^{+}(x,x')\}$ &  $-\f{\eta\eta'(\eta+\eta')(\eta+3\eta')}{36\,\alpha_{d}^{4}\,\pi^{4}(\eta-\eta')^{8}}$ \\ 
 \hline
 $-\f{2}{9\eta'}\{\p_{\eta}G^{+}(x,x')\}\{\p_{x}\,\p_{x'}G^{+}(x,x')\}\}$ &  $-\f{\eta\eta'(\eta+\eta')}{36\,\alpha_{d}^{4}\,\pi^{4}(\eta-\eta')^{7}}$ \\
 \hline 
 $-\f{2}{9}\{\p_{\eta'}\{G^{+}(x,x')\}\,\{\p_{\eta}\p_{x}\p_{x'}G^{+}(x,x')\}$ & $-\f{\eta\eta'(\eta+\eta')(3\eta+\eta')}{36\,\alpha_{d}^{4}\,\pi^{4}(\eta-\eta')^{8}}$\\
 \hline
 $+\f{1}{9}\{G^{+}(x,x')\}\{\p_{\eta}\p_{x}\p_{\eta'}\p_{x'}G^{+}(x,x')\}$ & $\f{\eta\eta(3\eta^{2}+14\eta\eta'+3\eta'^{2})}{72\,\alpha_{d}^{4}\,\pi^{4}(\eta-\eta')^{8}}$\\
  \hline
  $\f{1}{9\eta'}\{G^{+}(x,x')\}\{\p_{\eta}\p_{x}\p_{x'}G^{+}(x,x')\}$ & $\f{\eta\eta'(3\eta+\eta')}{72\,\alpha_{d}^{4}\,\pi^{4}(\eta-\eta')^{7}}$\\
  \hline
  $-\f{2}{9\eta}\{\p_{\eta'}G^{+}(x,x')\}\{\p_{x}\p_{x'}G^{+}(x,x')\}$ & $\f{\eta\eta'(\eta+\eta')}{36\,\alpha_{d}^{4}\,\pi^{4}(\eta-\eta')^{7}}$\\
  \hline
  $\f{1}{9\eta}\{G^{+}(x,x')\}\{\p_{x}\p_{\eta'}\p_{x'}G^{+}(x,x')\}$ & $-\f{\eta\eta'(3\eta+\eta')}{72\,\alpha_{d}^{4}\,\pi^{4}(\eta-\eta')^{7}}$\\
  \hline
  $\f{1}{9\eta\eta'}\{G^{+}(x,x')\}\{\p_{x}\p_{x'}G^{+}(x,x')\}$ & $-\f{\eta\eta'}{72\,\alpha_{d}^{4}\,\pi^{4}(\eta-\eta')^{6}}$\\
  \hline
\end{tabular}}
\end{center}
After adding all the terms we obtain the expression within the square bracket of Eq.(\ref{deriv_stress}). Now performing the similar analysis as the two-dimensional FLRW case, we land up to eq. (\ref{correlation_flw_5}). 

In case of the two-dimensionalSchwarzschild BH spacetime one can proceed similarly as the FLRW spacetime and in turn perceive the appearance of the eq. (\ref{correlation_sch_k1}).
\section{Schwinger function}\label{Schwinger}
The (1+1) Dimension FRW Universe result can be obtained explicitly from the Schwinger function also where the most general form of it is given by \cite{Francesco} 
\begin{eqnarray}
S_{abcd}(\mathbf{x}_{1},\mathbf{x}_{2})&&=\langle T_{ab}(\mathbf{x}_{1})T_{cd}(\mathbf{x}_{2})\rangle\nonumber\\
&&=\frac{A}{(\Delta\mathbf{x}^{2})^{4}}[(3\:g_{ab}g_{cd} - g_{ac}g_{bd} - g_{ad}g_{bc})(\Delta\mathbf{x}^{2})^{2}
- 4\Delta\mathbf{x}^{2}(g_{ab}\Delta x_{c}\Delta x_{d} + g_{cd}\Delta x_{a} \Delta x_{b})\nonumber 
\\
&&+8 \Delta x_{a}\Delta x_{b}\Delta x_{c}\Delta x_{d}]~,
\end{eqnarray}
where $\Delta\mathbf{x}^{2}= -(x^{0}-x^{0}\:')^{2}+ (x^{1}-x^{1}\:')^{2}$ and $A$ is an arbitrary constant, related to the central charge $C$ of the particular fields by $A=C/4\pi^{2}$. Since for the present case we have considered only massless scalar fields, its value is given by $A=1/4\pi^{2}$ as $C=1$. Using that technique one can obtain the form of stress tensor correlation function in (1+1) dimensions as
\begin{eqnarray}
&&a(t)a(t')\langle{0}\vert T^{tx}(t,x)T^{tx}(t',x')\vert{0}\rangle =\frac{1}{2^{6}\:\alpha_{d}^{4}\:\pi^{2}}\times\frac{1}{\sinh^{4}{\left(\frac{\Delta \tau}{2\:\alpha_{d}}\right)}}~,
\end{eqnarray}
in the proper frame of the FRW observer.
Similarly for BH result (\ref{correlation_sch_k}) can be obtained from the above Schwinger function.
\section{Table of Notations}
\begin{center}
 \scalebox{1.0}{\begin{tabular}{||c | c ||} 
 \hline
 \multicolumn{2}{|c|}{Vacuum States} \\
 \hline
 $|0\rangle$ & Conformal  \\ 
 \hline
 $|0\rangle_{M}$ & Minkowski  \\
 \hline
 $|0\rangle_{s}$ & Boulware  \\
 \hline
 $|0\rangle_{K}$ & Kruskal  \\
 \hline
 $|0\rangle_{U}$ & Unruh  \\ [1ex] 
 \hline
\end{tabular}}
\quad
 \scalebox{1.0}{\begin{tabular}{||c | c ||} 
 \hline
 \multicolumn{2}{|c|}{+ve frequency Wightman functions} \\
 \hline
 $G^{+}_{(2)}$ & $2D$ de Sitter Universe  \\ 
 \hline
 $G^{+}$ & $4D$ de Sitter Universe  \\
 \hline
 $G^{+}_{s}$ & $2D$ Boulware  \\
 \hline
 $G^{+}_{K}$ & $2D$ Kruskal  \\
 \hline
 $G^{+}_{U}$ & $2D$ Unruh  \\ [1ex] 
 \hline
\end{tabular}}
\end{center}

\begin{center}
\begin{tabular}{ ||c|c|| } 
 \hline
 $\mathcal{R}^{(2)}$ & Ricci Scalar in (1+1)$D$  \\ 
 $\mathcal{R}^{(4)}$ & Ricci Scalar in (3+1)$D$  \\ 
 $R$ & Constant Radial vector\\
 \hline
\end{tabular}
\end{center}


\end{document}